\documentclass{article}

\usepackage{arxiv}

\newtheorem{example}{Example}
\newtheorem{proposition}{Proposition}

\usepackage{hyperref}
\usepackage{amsmath}
\usepackage{graphicx}
\usepackage{caption}
\usepackage{subcaption}
\usepackage{tikz}
\usetikzlibrary{shapes, arrows.meta, positioning}

\begin{document}

\title{Biarchetype analysis: simultaneous learning of observations and features based on extremes}

\date{} 					

\author{Aleix~Alcacer \\
	Department of Mathematics\\
	Jaume I University\\
	\texttt{aalcacer@uji.es} \\
	\And
        Irene~Epifanio \\
	Department of Mathematics\\
	Jaume I University\\
	\texttt{epifanio@uji.es} \\
 	\And
        Ximo~Gual-Arnau \\
	Department of Mathematics\\
	Jaume I University\\
    \texttt{gual@uji.es}
}

\maketitle

\begin{abstract}
We introduce a novel exploratory technique, termed biarchetype analysis, which extends archetype analysis to simultaneously identify archetypes of both observations and features. This innovative unsupervised machine learning tool aims to represent observations and features through instances of pure types, or biarchetypes, which are easily interpretable as they embody mixtures of observations and features. Furthermore, the observations and features are expressed as mixtures of the biarchetypes, which makes the structure of the data easier to understand. We propose an algorithm to solve biarchetype analysis. Although clustering is not the primary aim of this technique, biarchetype analysis is demonstrated to offer significant advantages over biclustering methods, particularly in terms of interpretability. This is attributed to biarchetypes being extreme instances, in contrast to the centroids produced by biclustering, which inherently enhances human comprehension. The application of biarchetype analysis across various machine learning challenges underscores its value, and both the source code and examples are readily accessible in R and Python at \url{https://github.com/aleixalcacer/JA-BIAA}.

\end{abstract}


\section{Introduction}\label{sec:introduction}

Cluster analysis (CLA) is one of the most widely used tools in exploratory data analysis. The idea of clustering is to make groups of observations in such a way that each group contains similar observations that are different to those of the rest of the groups. If the data consist of well-separated clusters, appropriate clustering techniques can obtain, on the one hand, the representative of each cluster (the mean or centroid of the cluster for the popular $k$-means technique), and, on the other hand, the assignations of each observation  to one cluster, or a degree of belonging to each cluster for fuzzy clustering techniques. 

However, CLA is also used as a segmentation technique in the absence of well-separated (clearly differentiated) clusters in data. Many times, data follow a fan-spread pattern, i.e.  features vary continuously across observations.  The centroids are located in the middle of the data cloud since data points have to be covered in such a way that the distance between them and the assigned centroid is minimized (see \cite{7795738} about the relationship  between CLA and set  partitioning). In those cases, where data can be viewed as a superposition of various populations, it is of particular interest to use Archetype Analysis (AA) for segmenting  \cite{10.1007/978-3-030-33676-9_12}.

Instead of segmenting on the centroids, AA segments on the extremes. AA was defined by \cite{cutler1994archetypal}. The objective of AA is to represent the observations by means of a convex combination of archetypes, which in turn are convex combinations of observations. Archetypes or `pure types' lie on the boundary of the convex hull of the data and are therefore extreme profiles. Being extreme instances rather than central instances makes human understanding and interpretation of data easier \cite{Thurau12} since human cognition prefers extreme opposites \cite{Davis2010}. An illustrative example of this was analyzed in \cite{IsmaelTFM}, where CLA and AA were compared and archetypes were much more informative and understandable than centroids, because archetypes are further apart from each other than centroids.

Biclustering is a data mining technique introduced by \cite{doi:10.1080/01621459.1972.10481214}, although it was popularized by \cite{cheng2000biclustering}, who applied it to gene expression
data analysis. In biclustering (also known as block clustering, co-clustering, or two-mode clustering), rows (observations) and columns (features) of a data matrix are simultaneously clustered. An excellent overview of biclustering and fuzzy biclustering is found in \cite{FERRARO202163}. Biclustering is widely used in biological and medical applications \cite{zhao2012biclustering}, especially in gene expression data \cite{kerr2008techniques,10.1093/bib/bby014}. However, it is also applied in many other fields, such as marketing \cite{dolnicar2012biclustering}, psychology \cite{bib:van2004two},  recommender systems \cite{forsati2013fuzzy}, sports \cite{kaiser2011biclustering,shkedy2016identification}, website traffic \cite{koutsonikola2009fuzzy}, and many other pattern recognition applications, such as collaborative filtering, text mining, multimedia data processing and retrieval, etc. \cite{zhao2012biclustering,HENRIQUES20153941}.

In recent years, there has been growing interest in AA. On the one hand, there has been an increasing number of papers proposing efficient computational methods to calculate AA \cite{Morup2012,chen:hal-00995911,bauckhage2015archetypal,mair2017frame}, with applications in computer vision. On the other hand, AA has been applied in other very  diverse fields, such as, climatology \cite{doi:10.1175/JCLI-D-15-0340.1,w9110873},  ergonomics \cite{EpiVinAle,plos20}, genetics \cite{shoval2012evolutionary,Morup2013,wang2022non}, image processing  \cite{8237850,SUN2017147,DBLP:journals/remotesensing/SunZXTYL17,CabEpi19},  machine learning problems \cite{Morup2012,Rago15,math9070771,keller2021learning}, market research \cite{Porzio2008}, multi-document summarization \cite{Canhasi14}, nanotechnology \cite{doi:10.1021/acsnano.5b05788}, neuroscience \cite{griegos,Morup16}, sports \cite{Eugster2012,VinEpi17,VinEpi19} and sustainability \cite{Thurau12}. Finally, other papers have proposed extensions and new methodologies derived from AA with applications in  a broad
spectrum of fields: kernel AA \cite{Morup2012}, AA with missing data \cite{Morup2012,doi:10.1080/00031305.2018.1545700}, robust AA \cite{Eugster2010,Moliner2018a}, interval archetypes \cite{Esposito2012}, archetypoid analysis (ADA) \cite{Vinue15}, functional AA \cite{Epifanio2016}, data-driven prototype identification \cite{SAM16}, archetypal networks \cite{Rago15}, probabilistic AA \cite{Eugster14}, AA for nominal \cite{EugsterPAMI,IsmaelTFM} and ordinal  observations \cite{FERNANDEZ2021281}, directional AA \cite{10.3389/fnins.2022.911034}, AA for shapes \cite{EpiIbSi17}, deep AA \cite{van2019finding,keller2021learning}, and outlier detection \cite{MillanEpi,vinue2020robust,CABERO2021106830}. 
Nevertheless, no previous work has developed archetypal analysis for both rows and columns simultaneously, which we refer to as biarchetype analysis (biAA),  co-archetype analysis or two-mode archetype analysis.

\cite{10.1093/bib/bby014} reviews biclustering in biological and biomedical fields. They point out the need to improve the interpretability of biclustering results, and they also highlight that possible overlapping homogeneous submatrices have to be identified. This clashes with the idea of CLA, whose origin was to find separate (not overlapping) groups, but it is in the line with the basis of AA. Moreover, biclustering of human gene expression
data has been used to identify phenotype–genotype associations in studies of common or rare diseases. Note that archetypes themselves are phenotypes \cite{keller2021learning}; in fact, archetypes have been used also to explain  the evolutionary development of biological systems \cite{tendler2015evolutionary}. Therefore, putting all this together, it seems that biAA could be a reasonable alternative to biclustering in biology, as biAA could improve the interpretability of results. Nevertheless, the fields of application of biAA  are not just restricted to biology; they would be the same as for biclustering, i.e. biAA can be applied to many pattern recognition problems.

Our contributions consist of defining biAA for the first time, proposing a computational method to calculate it, whose implementation is available in the R package  biaa \href{https://github.com/aleixalcacer/biaa}{https://github.com/aleixalcacer/biaa} and the Python package \url{https://github.com/aleixalcacer/archetypes},  showing  how it works and the advantages of using archetypes (extremes) rather than the centroids of biclustering in an illustrative example, and finally, applying it to several real data sets in different fields to demonstrate the usefulness of biAA in various problems.

The outline of the paper is as follows: previous methodologies (CLA, biclustering, fuzzy biclustering, AA) are reviewed in a common framework in Sec. \ref{revision}. In Sec. \ref{metodo}, biAA is defined and a computational procedure is proposed. An illustrative example is used to exemplify biAA and compare it to biclustering.   In Sec. \ref{resultados}, our proposal is applied to three real data sets. Some conclusions and ideas for future work are provided in Sec. \ref{conclusiones}.

\section{Background \label{revision}}
Matrix factorization is our common framework for describing the established methods (as used in \cite{vichi2001} for clustering) and our proposal. Let ${\bf X}_{n \times m}$ be a data matrix with $n$ observations and  $m$ continuous features (they should be standardized in order to avoid problems if they measure different dimensions). Let ${\bf \alpha}_{n \times k}$ and ${\bf \gamma}_{c \times m}$ be matrices with values in $[0, 1]$. ${\bf \alpha}$ is the membership matrix of the observations, while ${\bf \gamma}$ is the membership matrix of the features. ${\bf Z}$ is the matrix of representative instances that approximates ${\bf X}$. The objective is to minimize: $\left \Arrowvert {\bf X} - {\bf \alpha Z \gamma}\right \Arrowvert^2$, with different constraints, where $\left \Arrowvert . \right \Arrowvert$ stands for the Frobenius norm. 

\subsection{Clustering}
For clustering, ${\bf Z}$ is the matrix of centroids, which is computed by ${\bf Z}$ = $({\bf \alpha}'{\bf \alpha})^{-1}{\bf \alpha}' {\bf X} {\bf \gamma}'({\bf \gamma}{\bf \gamma}')^{-1}$, where $'$ denotes transpose. 

{\bf{$k$-means clustering}}: The constraints are:  $\sum_{g=1}^k \alpha_{ig} = 1$ with $\alpha_{ig} \in \{0,1\}$ for $i=1,\ldots,n$ and  $\bf \gamma$ = ${\bf I}_{m \times m}$ is the identity matrix of order $m$. The matrix {${\bf Z}_{k \times m}$} = $({\bf \alpha}'{\bf \alpha})^{-1}{\bf \alpha}' {\bf X}$ has the centroids of each one of $k$ groups that partition the data set. 

{\bf{Fuzzy clustering}}: In soft clustering, each observation is assigned membership to each group.  The restrictions are:  $\sum_{g=1}^k \alpha_{ig} = 1$ with $\alpha_{ig} \geq 0$ for $i=1,\ldots,n$ and  $\bf \gamma$ = ${\bf I}_{m \times m}$. Again, the matrix {${\bf Z}_{k \times m}$} has the centroids of each one of $k$ groups. 

{\bf{Biclustering}}: This is also called double $k$-means with hard partitions by \cite{vichi2001}, where algorithms to solve it are proposed.  The constraints are: $\sum_{g=1}^k \alpha_{ig} = 1$ with $\alpha_{ig} \in \{0,1\}$ for $i=1,\ldots,n$  and  $\sum_{h=1}^c \gamma_{hj} = 1$ with $\gamma_{hj} \in \{0, 1\}$ for $j=1,\ldots,m$. Now, the dimension of ${\bf Z}$  is ${k \times c}$, since there are $k$ groups for observations and $c$ groups of variables.

{\bf{Fuzzy biclustering}}:  This is also called  fuzzy double $k$-means by \cite{vichi2001}. The constraints are now continuous: $\sum_{g=1}^k \alpha_{ig} = 1$ with $\alpha_{ig} \geq 0$ for $i=1,\ldots,n$  and  $\sum_{h=1}^c \gamma_{hj} = 1$ with $\gamma_{hj} \geq 0$ for $j=1,\ldots,m$. \cite{FERRARO202163} proposed several algorithms for solving fuzzy double $k$-means with continuous data, called FDkM and FDkMpf (Fuzzy Double $k$-Means with polynomial fuzzifiers), whose Matlab implementations are available in \cite{FERRARO202163}.

Besides the previous framework, there are some proposals of model-based biclustering. In that case, it is supposed that data are generated by a mixture distribution, as in \cite{GOVAERT2003463}, referred to as BMM (Block Mixture Model). Instead of memberships, it returns the final posterior probabilities for rows and columns, in addition to the mean and variance of each co-cluster. This is implemented in the R package  blockcluster \cite{JSSv076i09}. 

\subsection{Archetype analysis \label{revaa}}
In AA, $ {\bf Z}_{k \times m} = {\bf \beta}_{k \times n} {\bf X}_{n \times m} $, where $\sum_{l=1}^n \beta_{gl} = 1$ with $\beta_{gl} \geq 0$ for $g=1,\ldots,k$, i.e. the archetypes are mixture of the data. The other restrictions are: $\sum_{g=1}^k \alpha_{ig} = 1$ with $\alpha_{ig} \geq 0$ for $i=1,\ldots,n$ and $\bf \gamma$ = ${\bf I}_{m \times m}$. Therefore, the objective function to minimize subject to the previous constraints is:

\begin{equation} \label{eq:aa}
\begin{split}
    RSS = \left \Arrowvert {\bf X} - {\bf \alpha Z}\right \Arrowvert^2 = \left \Arrowvert {\bf X} - {\bf \alpha \beta X}\right \Arrowvert^2 = \\
    \sum_{i=1}^n\sum_{j=1}^m \left( x_{ij} - \sum_{g=1}^k \alpha_{ig} z_{gj} \right)^2 = \\
     \sum_{i=1}^n\sum_{j=1}^m \left( x_{ij} - \sum_{g=1}^k \alpha_{ig} \left( \sum_{l=1}^n \beta_{gl} x_{lj}   \right) \right)^2.
\end{split}
\end{equation}

The ${\bf \alpha}$ coefficients determine how much each archetype contributes
to the approximation of each observation, i.e. $\alpha_{ig}$ is the weight of the archetype $g$ for the $i$-th observation. Archetypes are built as mixtures of observations weighted by  ${\bf \beta}$ coefficients.

If $k$ = 1, the archetype coincides with the mean, but with $k$ $>$ 1, the archetypes are located on the boundary of the
convex hull of the data \cite{cutler1994archetypal}. Archetypes are
not necessarily nested, so different $k$s may
reveal distinct structures of the data. Therefore, as happens in other unsupervised statistical learning procedures, the selection of the number $k$ of prototypes has to be determined. If we have prior knowledge of
the arrangement of the data, $k$ can be selected based on this. Otherwise, we can use a simple but effective heuristic method, the elbow criterion, which has been used elsewhere \cite{cutler1994archetypal,Eugster2009}.  The elbow criterion consists of displaying the RSS for different $k$ values and choosing the value $k$ as the position where the elbow is located. This method is also used in clustering. 

\cite{cutler1994archetypal} proposed an alternating minimizing algorithm to find the matrices ${\bf \alpha}$ and ${\bf \beta}$ that minimizes RSS. This consists of alternating between estimating the best  ${\bf \alpha}$ for given archetypes ${\bf Z}$, and the optimum archetypes ${\bf Z}$ for  given ${\bf \alpha}$. In each phase, convex least squares
problems have to be solved. They used a penalized version of the non-negative
least squares algorithm \cite{Lawson74}.

\subsection{{Illustrative example of fuzzy $k$-means versus AA}}

\begin{figure}[!ht]
\centering
\subfloat[{Fuzzy $k$-means clustering assignments.}]{\includegraphics[width=0.3\columnwidth]{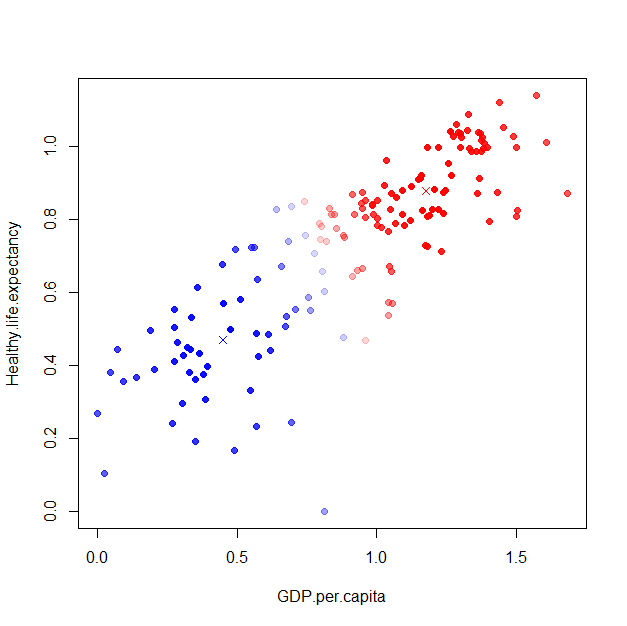}%
\label{fig:happyfc}}
\hfil
\subfloat[{AA assignments by the maximum $\alpha$.}]{\includegraphics[width=0.3\columnwidth]{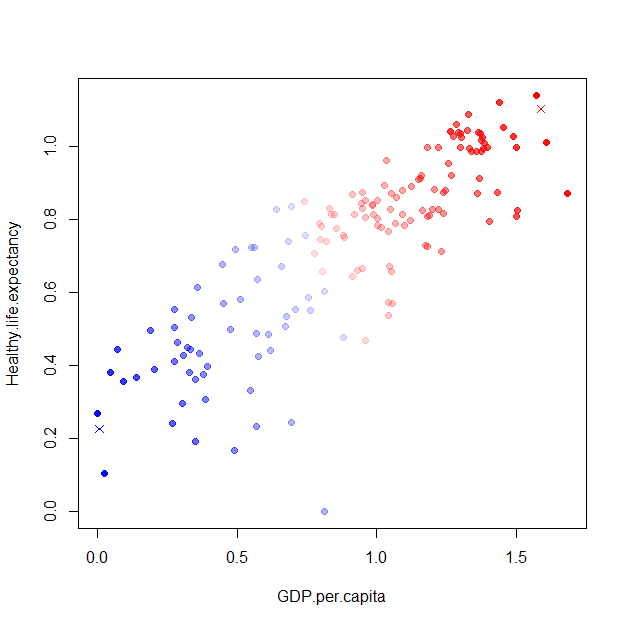}%
\label{fig:happyaa}}
\caption{{Plot of world happiness example. The crosses represent the prototypes.}}
\label{fig:happy}
\end{figure}

{
Fig. \ref{fig:happy} illustrates the outcomes of applying fuzzy \(k\)-means and AA with \(k = 2\) to variables associated with World Happiness, to demonstrate the concept of archetypes and their distinction from CLA. Contrary to CLA, AA models distinct aspects in the data rather than focus on the most central dynamics.  With AA, two archetypal countries are identified as prototypes. One is associated with Utopia, an imaginary nation characterized by the world's happiest populace, and the other with Dystopia, an imaginary nation marked by the world's least happy populace, commonly used in sociology as a benchmark. Countries are described as mixtures (encapsulated in $\alpha$ coefficients) of these idealized nations. This approach aligns with the human tendency to represent a group of objects by its extreme elements \cite{Davis2010}. However, with fuzzy \(k\)-means, the prototypes are situated in the middle of the data cloud; they are not the purest, hence their profiles are not as distinct as those of the archetypes. AA qualitatively offers a better explanation of the data structure. For instance, a country with values 0.657 and 0.672 for GDP.per.capita and Healthy.life.expectancy, respectively, is explained by AA as 43\% Utopia and 57\% Dystopia. Conversely, with CLA, this country falls into the blue cluster with a membership degree of 79\% and a centroid distance of 0.08. These results are paralleled by a country with values 0.268 and 0.242 for GDP.per.capita and Healthy.life.expectancy, which also has a centroid distance of 0.08 and a membership degree of 94\% in CLA. Yet, AA explains this country as 13\% Utopia and 87\% Dystopia. Therefore, AA  better distinguishes the difference in the profile’s  of these two countries.

}

\section{Biarchetype analysis \label{metodo}}
\subsection{Definition} \label{definicion}
In biAA, biarchetypes  are ${\bf Z}_{k \times c} = {\bf \beta}_{k \times n} {\bf X}_{n \times m} {\bf \theta}_{m \times c} $, where  $\sum_{l=1}^n \beta_{gl} = 1$ with $\beta_{gl} \geq 0$ for $g=1,\ldots,k$ and $\sum_{r=1}^m \theta_{rh} = 1$ with $\theta_{rh} \geq 0$ for $h=1,\ldots,c$, i.e. the archetypes are mixture of the data points and variables. There are $k$ archetypes for rows and $c$ for columns.  The other restrictions are: $\sum_{g=1}^k \alpha_{ig} = 1$ with $\alpha_{ig} \geq 0$ for $i=1,\ldots,n$ and $\sum_{h=1}^c \gamma_{hj} = 1$ with $\gamma_{hj} \geq 0$ for $j=1,\ldots,m$. Therefore, the objective function to minimize subject to the previous constraints is:

\begin{equation} \label{eq:biaa}
\begin{split}
    RSS =  \left \Arrowvert {\bf X} - {\bf \alpha Z \gamma}\right \Arrowvert^2 = \left \Arrowvert {\bf X} - {\bf \alpha \beta X \theta \gamma}\right \Arrowvert^2 = \\
    \sum_{i=1}^n\sum_{j=1}^m \left( x_{ij} - \sum_{g=1}^k\sum_{h=1}^c \alpha_{ig} z_{gh}\gamma_{hj} \right)^2 = \\
    \sum_{i=1}^n\sum_{j=1}^m \left( x_{ij} - \sum_{g=1}^k\sum_{h=1}^c \alpha_{ig} \left( \sum_{l=1}^n \sum_{r=1}^m \beta_{gl} x_{lr} \theta_{rh}  \right)\gamma_{hj} \right)^2.
\end{split}
\end{equation}

As before, the ${\bf \alpha}$ coefficients determine how much each archetype contributes
to the approximation of each observation, i.e. $\alpha_{ig}$ is the weight of the archetype $g$ for the $i$-th observation. Analogously, the ${\bf \gamma}$ coefficients determine how much each archetype contributes
to the approximation of each variable, i.e. $\gamma_{hj}$ is the weight of the archetype $h$ for the $j$-th variable. Biarchetypes are built as mixtures of observations and variables weighted by  ${\bf \beta}$ and ${\bf \theta}$ coefficients, respectively.

{

\subsection{Relationship with clustering}

Although the main objective of biarchetype analysis is to identify extreme values that define the dataset, it is worth noting that biAA can also be applied to clustering tasks, despite this not being its primary focus. 

Fig. \ref{scheme} displays a scheme showing the relationship between biAA and other unsupervised methods, where  $\sum_{g=1}^k \alpha_{ig} = 1$ with $\alpha_{ig} \geq 0$ for $i=1,\ldots,n$. biAA is to fuzzy biclustering as AA is to fuzzy clustering; and biAA is to AA as fuzzy biclustering is to fuzzy clustering. In simple words, in clustering methods, ${\bf Z}$ = $({\bf \alpha}'{\bf \alpha})^{-1}{\bf \alpha}' {\bf X} {\bf \gamma}'({\bf \gamma}{\bf \gamma}')^{-1}$ are centroids, but in archetypal methods, ${\bf Z} = {\bf \beta} {\bf X} {\bf \theta}$ are archetypes (extremes) ($\bf \theta$ = ${\bf I}_{m \times m}$ for AA);  only observations are considered in AA and fuzzy clustering; therefore,  $\bf \gamma$ = ${\bf I}_{m \times m}$, unlike biAA and fuzzy biclustering, where observations and variables are considered simultaneously.

}

{
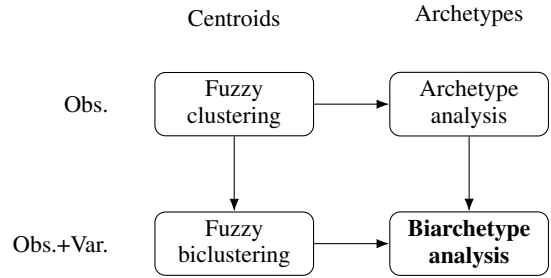
\begin{figure}[h!]
\centering
\begin{tikzpicture}[
    block/.style={rectangle, draw, text width=6em, text centered, rounded corners},
    line/.style={draw, -Latex}
]

\footnotesize
\node[block] (fuzzyclustering) {Fuzzy clustering};
\node[block, below=of fuzzyclustering] (fuzzybiclustering) {Fuzzy biclustering};
\node[block, right=of fuzzyclustering] (archetypeanalysis) {Archetype analysis};
\node[block, right=of fuzzybiclustering] (biarchetypeanalysis) {\textbf{Biarchetype analysis}};

\node[left=0.5cm of fuzzyclustering] (obs) {Obs.};
\node[left=0.5cm of fuzzybiclustering] (obsv) {Obs.+Var.};
\node[above=0.5cm of fuzzyclustering] (centroids) {Centroids};
\node[above=0.5cm of archetypeanalysis] (archetypes) {Archetypes};

\draw[line] (fuzzyclustering) -- (archetypeanalysis);
\draw[line] (fuzzybiclustering) -- (biarchetypeanalysis);
\draw[line] (fuzzyclustering) -- (fuzzybiclustering);
\draw[line] (archetypeanalysis) -- (biarchetypeanalysis);

\end{tikzpicture}
\caption{{Diagram of the relationship between biAA and other unsupervised methods.}}
\label{scheme}
\end{figure} 
}

\subsubsection{Location of biarchetypes}
In this section, we state some results that help in understanding the behavior of the row and column vectors of the biarchetype matrix ${\bf Z}_{k \times c}$.
\medskip

Let ${\bf X}_{n \times m}$ be a data matrix with $n$ observations and  $m$ continuous features. We denote by ${\bf x}_i^d$, $i=1,\dots,n$ the row vectors of the matrix ${\bf X}_{n \times m}$ (in this case observations) and by ${\bf x}_i^f$, $i=1,\dots,m$ the column vectors of the matrix ${\bf X}_{n \times m}$ (in this case features). This notation will be the same for all the matrices used.

The problem is to find a matrix ${\bf Z}_{k \times c}$ with $1\le k\le n$, $1\le c\le m$, which is expressed as $ {\bf Z}_{k \times c} = {\bf \beta}_{k \times n} {\bf X}_{n \times m} {\bf \theta}_{m \times c}$ and the matrices $\alpha_{n\times k}$, ${\bf Z}_{k \times c}$ and $\gamma_{c\times m}$ minimize 
$$ RSS =  \left \Arrowvert {\bf X}_{n \times m} -  \alpha_{n \times k} {\bf Z}_{k \times c} \gamma_{c \times m}\right \Arrowvert^2.$$

Now we distinguish several cases depending on the values of $k$ and $c$.

{\bf Case I: $k=1$ and $c=1$}. In this case, $\alpha_{n\times 1}=(1,\dots,1)'$  and $\gamma_{1\times m}=(1,\dots,1)$; then, the real value ${\bf Z}_{1 \times 1}$ that minimizes RSS is the mean of all entries in matrix ${\bf X}_{n \times m}$, that is,
$${\bf Z}_{1 \times 1} = \frac{1}{nm}\sum_{i=1}^n\sum_{j=1}^m x_{ij}.$$
\medskip

{\bf Case II: $k=n$ and $c=m$}. In this case we consider $\alpha_{n\times n}={\bf I}_{n\times n}$ and $\gamma_{m\times m}= {\bf I}_{m\times m}$. Then, by choosing ${\bf Z}_{n \times m}= {\bf X}_{n \times m}$ we obtain RSS = 0. 
\medskip

{\bf Case III: $1\le k < n$ and $c=m$}. We consider $\gamma_{m\times m}= {\bf I}_{m\times m}$ and $\theta_{m\times m}= {\bf I}_{m\times m}$, then, the problem now consists of minimizing $$ RSS =   \left \Arrowvert {\bf X}_{n \times m} -  \alpha_{n \times k} {\bf Z}_{k \times m} \right \Arrowvert^2,$$
which is a typical problem in AA and the location of the archetypes is explained in  \cite{cutler1994archetypal} and reviewed in Sec. \ref{revaa}.
\medskip

{\bf Case IV: $k= n$ and $1\le c < m$}. This is the case of finding only the archetypes of features, i.e. we consider ${\bf \alpha} = {\bf I}_{n \times n}$ and ${\bf \beta} = {\bf I}_{n \times n}$. As the Frobenius norm of a matrix is the same as the Frobenius norm of its transpose, $\left \Arrowvert {\bf X} - {\bf Z \gamma}\right \Arrowvert^2$ =
$\left \Arrowvert {\bf X}' - {\bf \gamma' Z'}\right \Arrowvert^2$, features can adopt the role of observations when ${\bf X}$ is transposed. Then, the same reasoning as in the preceding case entails a problem of AA. 
\medskip

{\bf Case V: $k= 1$ and $1< c < m$}. In this case ${\bf \alpha} = {\bf I}_{n \times n}$ and the problem of minimizing 
$$ RSS =   \left \Arrowvert {\bf X}_{n \times m} - (1,\dots,1)' \left( {\bf Z}_{1 \times c}  \gamma_{c\times m}\right) \right \Arrowvert^2,$$
is satisfied if the mean value  
$\frac1{n} \sum_{i=1}^n {\bf x}_i^d=(\bar{x}_1^d, \dots, \bar{x}_m^d) =\left( {\bf Z}_{1 \times c}  \gamma_{c\times m}\right)$ where $\bar{x}_j^d=\frac1{n}\sum_{i=1}^n x_{ij}$.  Note that each real number $\bar{x}_i^d$ belongs to the convex hull of the components of the vector ${\bf Z}_{1 \times c}$.
\medskip

{\bf Case VI: $c= 1$ and $1< k < n$}. In this case RSS is minimized if 
$\frac1{m} \sum_{i=1}^m {\bf x}_i^f=(\bar{x}_1^f, \dots, \bar{x}_n^f) =\left( {\bf Z}'_{k \times 1}  \alpha'_{n\times k}\right)$ where $\bar{x}_i^f=\frac1{m}\sum_{j=1}^m x_{ij}$.  Note that each real number $\bar{x}_i^f$ belongs to the convex hull of the components of the vector ${\bf Z}_{k \times 1}$.
\medskip

{\bf Case VII: $1< k < n$ and $1<c<m$}. 

Let us call $ {\bf V}_{n \times c} =  {\bf X}_{n \times m} {\bf \theta}_{m \times c}$; then, each ${\bf v}_j^f$ $(j=1,\dots, c)$ belongs to the convex hull $C_X^f$ of the data  ${\bf x}_i^f$ $(i=1,\dots, m)$. Moreover, since $ {\bf Z}_{k \times c} = {\bf \beta}_{k \times n} {\bf V}_{n \times c}$, each vector ${\bf z}_j^d$ $(j=1,\dots, k)$ belongs to the convex hull $C_V^d$ of the vectors  ${\bf v}_i^d$ $(i=1,\dots, n)$.
\medskip

\begin{proposition} Having fixed the matrix ${\bf \theta}_{m \times c}$, there is a matrix of biarchetypes $ {\bf Z}_{k \times c}$ such that each row vector ${\bf z}_j^d$ $(j=1,\dots, k)$ belongs to the boundary of the convex hull $C_V^d$.
\end{proposition}
\medskip

Now, let us call $ {\bf Y}_{k \times m} =  \beta_{k \times n} {\bf X}_{n \times m}$; then, each ${\bf y}_j^d$ $(j=1,\dots, k)$ belongs to the convex hull $C_X^d$ of the data  ${\bf x}_i^d$ $(i=1,\dots, n)$. Moreover, since $ {\bf Z}_{k \times c} = {\bf Y}_{k \times m} {\bf \theta}_{m \times c}$, each vector ${\bf z}_j^f$ $(j=1,\dots, c)$ belongs to the convex hull $C_Y^f$ of the vectors  ${\bf y}_i^f$ $(i=1,\dots, m)$.
\medskip

\begin{proposition} Having fixed the matrix ${\bf \beta}_{k \times n}$, there is a matrix of biarchetypes ${\bf Z}_{k \times c}$ such that each column vector ${\bf z}_j^f$ $(j=1,\dots, c)$ belongs to the boundary of the convex hull $C_Y^f$.
\end{proposition}
\medskip

\noindent Proof of Propositions 1 and 2 is detailed in Appendix \ref{ape}. 

\bigskip

\begin{example} This toy example illustrates the location of the biarchetypes for different values of $k$ and $c$, for the following matrix $\begin{pmatrix}
1 & 2 & 3 & 4 & 5\\
6 & 7 & 8 & 9 & 10 \\
11 & 12 & 13 & 14 & 15\\
16 & 17 & 18 & 19 & 20\\
21 & 22 & 23 & 24 & 25\\
\end{pmatrix}$.
\medskip

For $k$ = 1 and $c$ = 1, $\bf z$ = 13 (mean of all entries of the matrix according to  Case I). For $k$ = 1 and $c$ =2, $\bf z$ = $\begin{pmatrix} 11 & 15\end{pmatrix}$; here $(\bar{x}_1^d, \dots, \bar{x}_5^d)=(11,12,13,14,15)$ and each real number $\bar{x}_i^d$ belongs to the convex hull of the components of $\bf z$ according to Case V. For $k$ = 2 and $c$ = 1, $\bf z$ = $\begin{pmatrix} 3\\ 23 \end{pmatrix}$ and, according to Case VI, each real number $\bar{x}_i^f$ belongs to the convex hull of the components of the vector $\bf z$. For $k$ = 2 and $c$ = 2, $\bf z$ = $\begin{pmatrix} 1 & 5\\21 & 25 \end{pmatrix}$. For this last case, RSS = 0, and, according to Case VII, the vectors ${\bf z}_j^d$ and ${\bf z}_j^f$, $j=1,2$ are located at the boundary of convex sets.

\end{example}

\begin{example} \label{ex2}
In this example, we will generate the data from a multivariate random distribution.

The covariance matrix for both rows and columns is:
\begin{align*}
\scriptsize{
\Sigma = \begin{pmatrix}
1 & 0.8 & \dots & 0.8 & 0.8 & 0 & 0 & \dots & 0 & 0 \\
0.8 & 1 & \dots & 0.8 & 0.8 & 0 & 0 & \dots & 0 & 0 \\
\vdots & \vdots &\ddots & \vdots & \vdots & \vdots & \vdots & \ddots & \vdots & \vdots \\
0.8 & 0.8 & \dots & 1 & 0.8 & 0 & 0 & \cdots & 0 & 0 \\
0.8 & 0.8 & \dots & 0.8 & 1 & 0 & 0 & \cdots & 0 & 0 \\
0 & 0 & \dots & 0 & 0  & 1 & 0.8 & \cdots & 0.8 & 0.8 \\
0 & 0 & \dots & 0 & 0 & 0.8 & 1 & \cdots & 0.8 & 0.8\\
\vdots & \vdots &\ddots & \vdots & \vdots & \vdots & \vdots & \ddots & \vdots & \vdots \\
0 & 0 & \dots & 0 & 0 & 0.8 & 0.8 & \cdots & 1 & 0.8 \\
0 & 0 & \dots & 0 & 0 & 0.8 & 0.8 & \cdots & 0.8 & 1 \\
\end{pmatrix}
}
\end{align*}

And the mean also for  both rows and columns is:
\begin{align*}
    \mu = (0 \cdots 0)
\end{align*}

\begin{figure}[!ht]
\centering
\subfloat[Simulated data as a heatmap.]{\includegraphics[width=0.4\columnwidth]{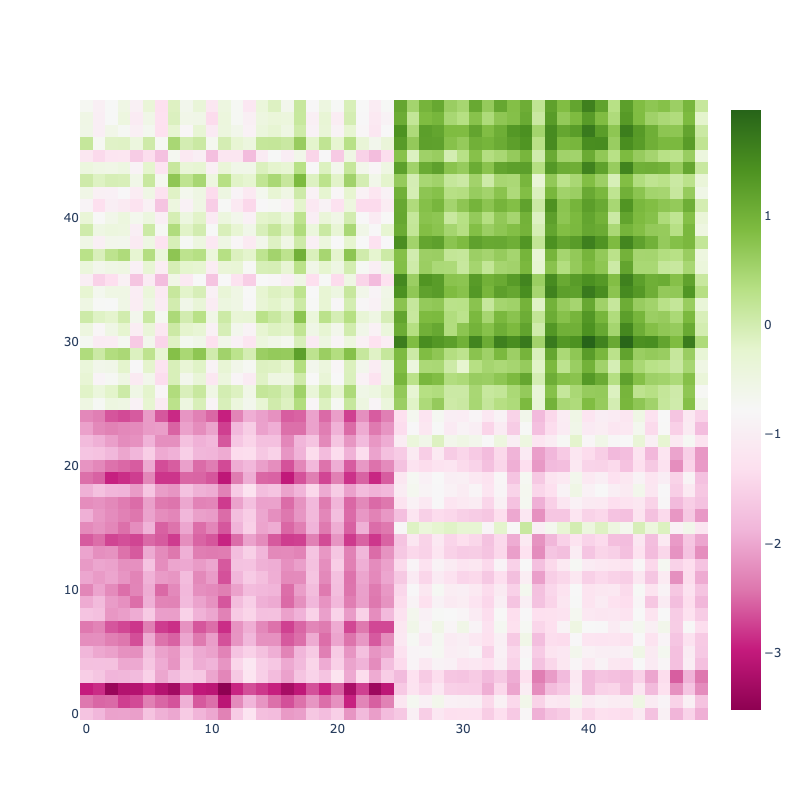}%
\label{fig:sim:data}}
\hfil
\subfloat[Simulated data in the biarchetype space.]{\includegraphics[width=0.4\columnwidth]{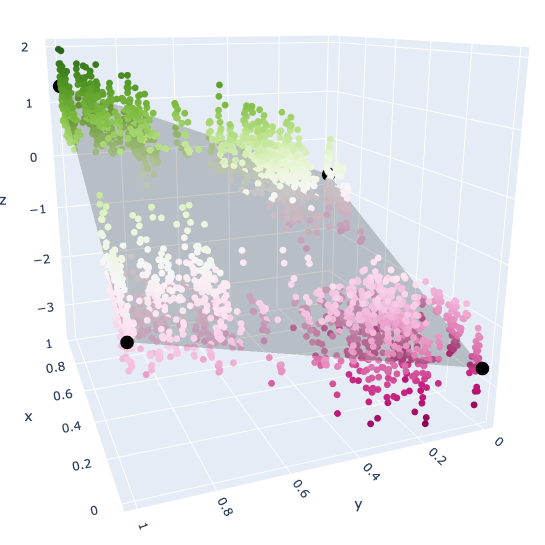}%
\label{fig:sim:arch}}
\caption{Representations of data and the biarchetypal space for example \ref{ex2}, i.e. representation of the coefficients $\alpha$ and $\gamma$.}
\label{fig:sim}
\end{figure}

The data generated can be seen in Fig. \ref{fig:sim:data}. In Fig. \ref{fig:sim:arch}, the data are represented using the coefficients $\alpha$ and $\gamma$ of biAA with $k$ = 2 and $c$ = 2. These coefficients are mapped to the $x$ and $y$ axes of the figure and the $z$ axis represents the value of each observation. The biarchetypes are represented in black and, as seen, they are at the extremes of the data.

\end{example}

\subsubsection{Selecting the number of biarchetypes} As in the case of AA, if there is no  information available a priori, we can use the elbow criterion, but in this case, we look for the elbow of a surface instead of a curve. In biAA, we run biAA for different values of $k$ and $c$ and display their RSS values in a 3D plot. We select the point ($k$,$c$) where the surface ``flattens'', i.e. $(k,c)$ is the point at which the RSS of the following points $(k+1,c)$, $(k,c+1)$, and $(k+1, c+1)$ stops decreasing drastically with respect to the RSS of the point $(k, c)$.

\subsection{Algorithm} The following iterative method is proposed to solve biAA. It is based on alternating minimization as in the AA algorithm by \cite{cutler1994archetypal}.
\begin{enumerate}
    \item Data preparation: To randomly initialize  the matrices $\bf \alpha$, $\bf \gamma$, $\bf \beta$ and $\bf  \theta$, fulfilling the constraints in Sect. \ref{definicion}.
    
    \item Repeat until RSS is  sufficiently small or the number of maximum iterations is reached:
        \begin{enumerate}
            \item Find the best $\bf \alpha$ (fixed $ \bf \gamma$): solve $n$  convex least squares problems using ${\bf X}' = ({\bf Z \gamma})' {\bf \alpha}'$.
            \item Find the best $\bf \gamma$ (fixed $\bf \alpha$): solve $m$  convex least squares problems using $\bf X$ = $({\bf \alpha Z}) {\bf \gamma}$.
            \item Recalculate the biarchetypes, where $^+$ stands for the Moore-Penrose pseudoinverse: ${\bf Z} = {\bf \alpha}^+ {\bf X \gamma}^+$.
            \item Find the best $\bf \beta$  (fixed $\bf \theta$): solve $k$  convex least squares problems using ${\bf Z}' = ({\bf X \theta})' {\bf \beta}'$.
            \item Find the best $\bf \theta$ (fixed $\bf \beta$): solve $c$  convex least squares problems using ${\bf Z} = ({\bf \beta X}) {\bf \theta}$.
            \item Recalculate the biarchetypes: ${\bf Z} = {\bf \beta X \theta}$.
            \item Calculate the new RSS.

        \end{enumerate}
\end{enumerate}

Note that the convex least squares problems can be solved as proposed by \cite{cutler1994archetypal}, i.e. using a penalized least squares problem \cite{Lawson74}. The idea is, given a least squares problem ${\bf A}_{n\times k} {\bf X}_{k\times m} = {\bf B}_{n \times m}$,  to add a row with constant elements $C$ to $\bf A$ and $\bf B$, in order to obtain a new problem ${\bf A}_{(n + 1) \times k} {\bf X}_{k\times m} = {\bf B}_{(n+1) \times m}$, in such a way that RSS would be:

\begin{equation}\label{convex}
\begin{aligned}
\text{\footnotesize $RSS =  \sum_{i=1}^{n+1}\sum_{j=1}^m \left (b_{ij} - \sum_{h=1}^k a_{ih}x_{hj} \right)^2 =$ } \\ 
  \text{\footnotesize $\sum_{j=1}^m \left( \sum_{i=1}^n \left (b_{ij} - \sum_{h=1}^k a_{ih}x_{hj} \right)^2 +  \left ( b_{n+1,j} - \sum_{h=1}^k a_{n+1,h}x_{hj} \right)^2 \right) =$} \\
 \text{\footnotesize $\sum_{j=1}^m \left( \sum_{i=1}^n \left (b_{ij} - \sum_{h=1}^k a_{ih}x_{hj} \right)^2 + \sum_{j=1}^m \left ( C - \sum_{h=1}^k Cx_{hj} \right)^2 \right) =$} \\
 \text{\footnotesize   $\sum_{j=1}^m \left(\sum_{i=1}^n \left (b_{ij} - \sum_{h=1}^k a_{ih}x_{hj} \right)^2 + \sum_{j=1}^m C^2\left ( 1 - \sum_{h=1}^k x_{hj} \right )^2\right)$}. \\
\end{aligned}
\end{equation}

Therefore, if value $C$  is high, the term $C^2\left ( 1 - \sum_{h=1}^k x_{hj} \right )^2$ forces the convexity of the elements of $\bf X$ in eq. \ref{convex}.

{As regards computational complexity, the biAA algorithm can be considered as complex as computing the AA algorithm twice. Like AA, the  speed of biAA depends on the efficiency of the convex least squares method. The computational complexity  for the AA algorithm  was analyzed by \cite{Eugster2009}, based on this analysis, the computation time increases linearly as the number of observations increases, while it remains approximately constant as the number of archetypes increases. In practical terms, the biAA algorithm is a computer-intensive algorithm and its convergence speed depends on the data structure, so if convergence is not attained in a few steps for  specific numbers $k$ and $c$,  those numbers probably do not explain the data well.}

\subsection{Illustrative example and comparison with biclustering \label{ilu}} 
\begin{figure}[!t]
\centering
\includegraphics[width=0.4\textwidth]{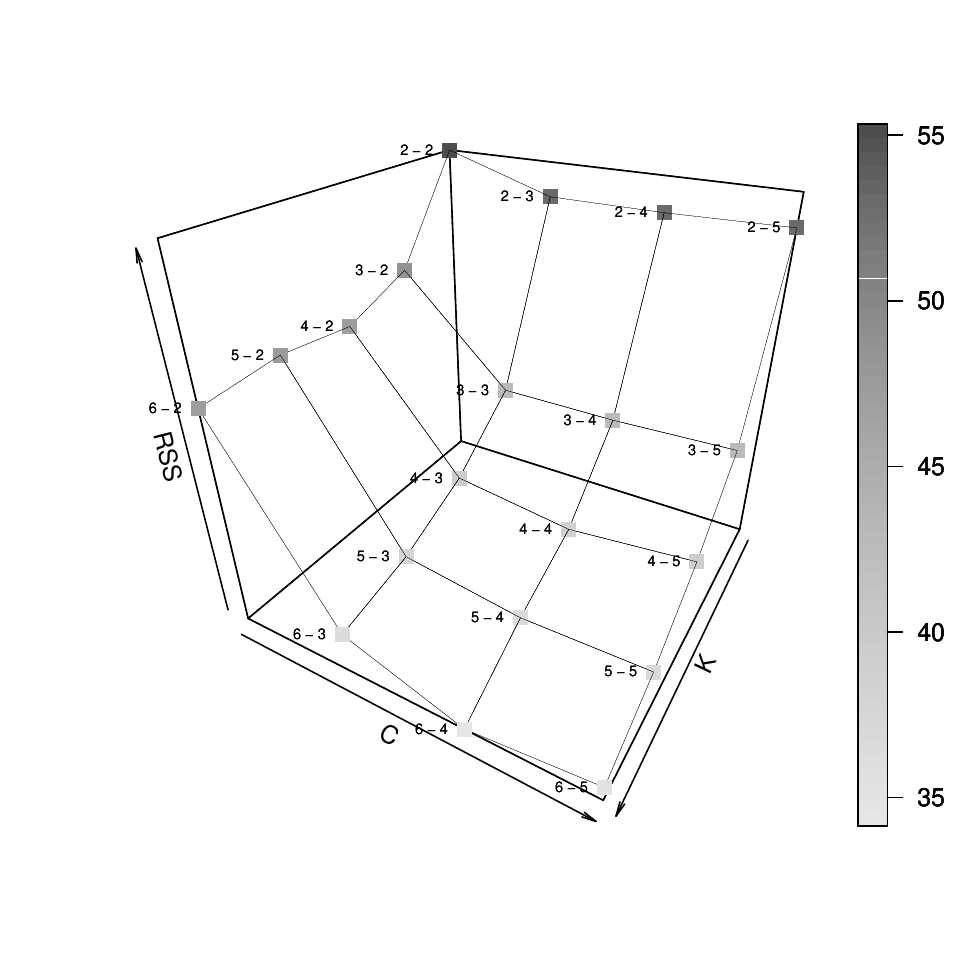}
\caption{RSS for $k$ from 2 to 6 and $c$ from 2 to 5.}
\label{rsscargas}
\end{figure}

The following example illustrates the use of biAA and its advantages in comparison with biclustering, especially when working with non-clustered data. We consider the data of 45 students from  Universitat Jaume I, who reported the number of hours per week spent working on a subject at home over 17 weeks. The complete description of the data can be found in \cite{epi2016cargas}. Missing data are imputed by $mice$ \cite{JSSv045i03}. The data range from 0 to 10, with mean 4.94 and standard deviation 2.46. We  apply biAA with $k$ = 4 and $c$ = 3, since the elbow is found at those values (see Fig. \ref{rsscargas}).

The $\gamma$ coefficients for biAA, BMM, and FDkMpf  are shown in Table \ref{cargasgamma} (we sort them in order to make the comments easier). FDkM  was also applied, but the solution is not valid since the same prototypes are obtained for all the groups.

\begin{table}[!t]
\centering
\caption{The $\gamma$ coefficients for biAA, BMM, and FDkMpf of the illustrative example.}
\label{cargasgamma}
\begin{tabular}{|ccc|ccc|ccc|}
\hline
1&0&0&1&0&0&1&0&0\\
0.71&0.12&0.17&1&0&0&1&0&0\\
0.57&0.42&0.01&1&0&0&0.85& 0.08&0.08\\
0&0.80&0.20&0&0.01&0.99&0&0.5&0.5\\
0.13&0.76&0.11&0&1&0&0&0.5&0.5\\
0.14&0.86&0&0&1&0&0.29&0.35&0.35\\
0&0.80&0.20&0&1&0&0&0.5&0.5\\
0.01&0.99&0&0&0&1&0.2&0.4&0.4\\
0&0.90&0.11&0&1&0&0&0.5&0.5\\
0&0.89&0.11&0&1&0&0&0.5&0.5\\
0&0.42&0.58&0&1&0&0&0.5&0.5\\
0.05&0.45&0.5&0&0.96&0.04&0&0.5&0.5\\
0&0.62&0.38&0&1&0&0&0.5&0.5\\
0.05&0&0.95&0&0&1&0&0.5&0.5\\
0&0.04&0.96&0&0&1&0&0.5&0.5\\
0&0.10&0.91&0&0&1&0&0.5&0.5\\
0.13&0&0.87&0&0&1&0&0.5&0.5\\
\hline
\end{tabular}
\end{table}

(From now on, we will use archetype or archetypal instead of biarchetype or biarchetypal to simplify the language).

The feature similar to the first archetypal variable corresponds to the first week. The second and third weeks are also similar, but with a temporal gradation (0.71 and 0.57). Week 8, an intermediate week of the semester, is similar to the second archetypal variable. Other intermediate weeks (4, 5, 6, 7, 9, and 10) are also similar. Week 15, a week at the end of the semester, is similar to the third archetypal variable, as well as  weeks 14, 16, and 17. Weeks 11, 12, and 13 are explained as mixtures (nearly 50\% -50\%) of the second and third archetypal variables. Note that the third week was also explained as a mixture close to 50\% -50\% of the first and second archetypal variables. In summary, the archetypal variables correspond to the profile of the beginning, middle and end of the semester, respectively. The weeks in the transitions between these temporal points are reflected as mixtures.

As regards the prototypical variables for biclustering methodologies, the first three weeks have probabilities of 1 (or nearly 1 for FDkMpf) for the first prototypical variable of BMM. Unlike biAA where gradation was found,  the probabilities (memberships) are nearly crisp classifications for BMM, not only for the first prototypical variable, but for the rest as well. The intermediate weeks 5, 6, 7, 9, 10, 11, 12, and 13 have probabilities of 1 for the second prototypical variable, while the final weeks (14, 15, 16, and 17), and the intermediate weeks 4 and 8  have probabilities of nearly 1 or 1  for the third prototypical variable. 

Note the difference with biAA. On the one hand, in BMM there is a lack of gradation over time in the memberships (no mixture is found, but the memberships are extremely high, nearly all ones), as if changes between adjoining weeks were radical (as breaking jumps) rather than smooth. Therefore, the information provided by biAA  is richer. On the other hand, there are two intermediate weeks (4 and 8) belonging to the third prototypical variable corresponding to the end of semester weeks, which is not very coherent. Therefore, the information provided by biAA is more reasonable. Finally,  the second and third prototypical variables are identical for FDkMpf, with  50\%-50\% or close degrees of membership for weeks 4 to 17. Therefore, the information returned by FDkMpf is poorer than that of BMM and biAA.

Table \ref{Zcargas} displays the representative points ${\bf Z}$, archetypes or centroids for biAA and biclustering, respectively (we sort them in order to make the comments easier). The first archetype describes a student who works very few  hours per week throughout the semester. The second archetype represents a student who studies very few hours throughout the semester, except at the end of the semester, when they work for 9 hours per week. The third archetype describes a student who works very few hours at the beginning of the semester (1h per week), many hours during the semester (10h per week), and intermediate hours (4h per week) at the end of the semester. The fourth archetype represents a student who studies many hours throughout  the whole semester. 

For BMM, the prototypes are not as pure as the archetypes. For example, there is no great difference between centroids 2 and 3:  centroid 3 studies only one  or one  and a half hours more than centroid 2 per week. The centroids are not as intuitively interpretable as archetypes. Centroid 1 corresponds to a student who studies 2 or 3 hours throughout  the semester; centroid 2 studies 2h per week at the beginning and 4 or 5h per week for the rest of the semester; centroid 3 works 4 hours at the beginning and 6 hours per week for the rest of the semester; while centroid 4 studies 5h per week at the beginning of the semester and 7 or 8 hours throughout  the rest of the semester. It seems that centroids are limited to following a gradation according to the total number of hours studied throughout  the semester rather than by  differences in behavior throughout  the semester. For FDkMpf, the comments are similar, but in addition there is no difference between the intermediate and final weeks. For example, the profiles of  students 32 and 33, who are similar to archetype 2 (with $\alpha$s of 0.84 and 0.88, respectively), would not be reflected by the centroids of BMM or FDkMpf. They belong to cluster 2 of BMM, with probabilities of 0.97 and 1, respectively. But this does not say anything about how far (or in which direction) from centroid 2 those students are. This happens because the goal of clustering is to assign the data to groups, not to
explain the structure of the data more qualitatively. 

\begin{table}[!ht]
\centering
\caption{The archetypes and centroids for biAA, BMM, and FDkMpf of the toy example.}
\label{Zcargas}

\begin{tabular}{|ccc|ccc|ccc|}
\hline
1.54 & 2.36 & 0.36 &  1.67 & 2.07 & 3.49 & 2.03 & 3.28 & 3.28 \\
1 & 2 & 9 & 2.32 & 4.15 & 5.02 & 2.74 & 4.89 & 4.89\\
1 & 10 & 4 & 3.79 & 5.70 & 6.05 & 3.42 & 5.97 & 5.97\\
8 & 6.32 & 9.36 & 5.01 & 7.69 & 6.60 & 5.55 & 6.97 & 6.97\\
\hline
\end{tabular}
\end{table}

{
\subsection{Ablation study}

Another important aspect to consider is the ablation study. In this analysis, we study the implications of calculating archetypes separately (solely by rows and solely by columns) and then combining them, as opposed to computing the archetypes simultaneously.

In particular, the biarchetypes in biAA are reconstructed from the $\beta$s and $\theta$s obtained by applying ${\bf X} \simeq \alpha \beta {\bf X} \theta \gamma$ to the dataset ($arch_{biaa} = \beta  {\bf X}  \theta$), whereas those in the ensemble are derived from applying AA across the rows of ${\bf X}$, (i.e. ${\bf X} \simeq \alpha_r \beta_r {\bf X}$) and also applying it across the columns (i.e. $X \simeq {\bf X} \beta_c \alpha_c$). In the latter case, the biarchetypes are obtained by combining the $\beta$s from both methods ($arch_{ensemble} = \beta_r  {\bf X}  \beta_c$).

To conduct this study, we initiated an experimental procedure. We generated a synthetic dataset 50 times whose shape is $100\times 100$ and with $3\times 3$ biarchetypes. The values for these datasets were constrained to fall within the range of 0 to 1, ensuring a standardized scale for comparison.

Specifically, each dataset is constructed to conform a mixture of biarchetypes, where the ${\bf Z}$ matrix encapsulates the biarchetype values. Moreover, the entries of matrix $\alpha$'s rows and matrix $\gamma$'s columns are determined by sampling from the $U(0, \phi)$ distribution, with $\phi$ being a parameter within the interval [0, 1]. After that, the procedure assigns a 1 to the entry that denotes the group to which the observation belongs, thereby ensuring that this maximum coefficient signifies the group assignment. Subsequently, this vector is normalized to achieve a unit norm, thus ensuring its convexity.

The parameter $\phi$ significantly influences the membership characteristics within the archetypal model. At a setting of $\phi = 0$, each observation is completely represented by one archetype, exhibiting pure membership. As $\phi$ approaches 1, the model shifts towards a mixed-membership framework, allowing observations to have more evenly distributed memberships across the different archetypes. In this instance, the parameter $\phi$ has been determined to be 0.05.

\begin{figure}[!ht]
    \centering
    \includegraphics[width=0.7\columnwidth]{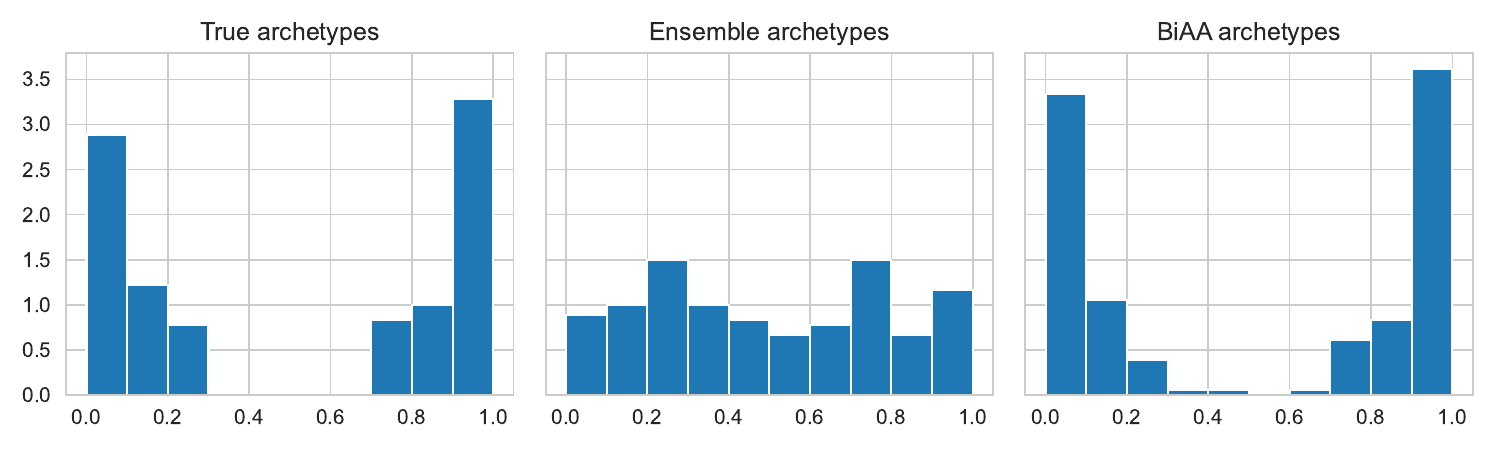}
    \caption{Distribution of the archetype values.}
    \label{fig:arch-dist}
\end{figure}

The histograms, as depicted in Figure~\ref{fig:arch-dist}, reveals a notable distinction in the distribution of values obtained using our methodology compared to the separate calculation of archetypes. Specifically, the values derived from the biAA approach manages to recover the true archetypes,  which does not happen with the ensemble archetype approach. 
}

\section{Results and discussion \label{resultados}}

Like biclustering, biAA can be applied to a wide range of fields. In this section, we will apply it to biology, document analysis and community detection. 

The code and data sets for reproducing the results including those in Sec. \ref{ilu} are available at \url{https://github.com/aleixalcacer/JA-BIAA}.

\subsection{Gene expression data}

To show how biAA can be applied, we examine data from gene expression of cutaneous melanoma used in \cite{Bittner2000, Rocci2008, MartellaAlfoVichi+2008}. Instead of meticulously re-analyzing this data set, we use it to highlight the salient features of biAA.

The aim of this study was to test the idea that molecular profiles generated by cDNA microarrays could be used to differentiate between several subtypes of cutaneous melanoma, a kind of skin cancer. mRNA was collected from the 31 cutaneous melanoma samples, and Cy5-labeled cDNA was created. All samples were examined with the same reference probe, identified as Cy3. For each sample, Cy5 and Cy3-labeled cDNA were combined and hybridized to a different melanoma microarray. Red and green lasers were used to scan the hybridization array, and the resulting image was then analyzed.

The same pre-processing was carried out as in \cite{Bittner2000, Rocci2008}. Only 3613 cDNAs of the 8150 observations  were classified as well measured. Cy5/Cy3 expression ratios were computed for the accurately measured genes. Ratios that were more than or equal to 50 and less than or equal to 0.02 were reduced to 50 and 0.02, respectively. A logarithmic scale was applied to the derived ratios (base 2). The log ratios were adjusted so that the median log-ratio for each experiment was equal to zero by subtracting the median log-ratio within an experiment from all log-ratios for that experiment. Since a single reference probe was utilized in all experiments, there was no standardization between trials.

Using one minus the Pearson correlation coefficient of log-ratios as a measure of dissimilarity between two experiments, \cite{Bittner2000} applied the average linkage hierarchical clustering on the 31 cutaneous melanoma samples and obtained two clusters of 12 and 19 samples.

Regarding \cite{Rocci2008}, they used double $k$-means to analyze the same data set. In this case, the columns were centered and scaled to unit variance, finding that the separation between two columns is proportional to one minus the Pearson correlation coefficient. In particular, their analysis indicates that samples 4 and 7 are members of the `19-samples' group obtained by \cite{Bittner2000}, i.e. the main cluster group.  The membership of these two samples in \cite{Rocci2008} differs from that obtained by \cite{Bittner2000}.

In our case, we applied biarchetype analysis to the same data set as \cite{Rocci2008}, extracting three archetypes for the genes (rows) and two archetypes for the melanoma samples (columns).

\begin{figure}[!t]
\centering
\includegraphics[width=0.5\columnwidth]{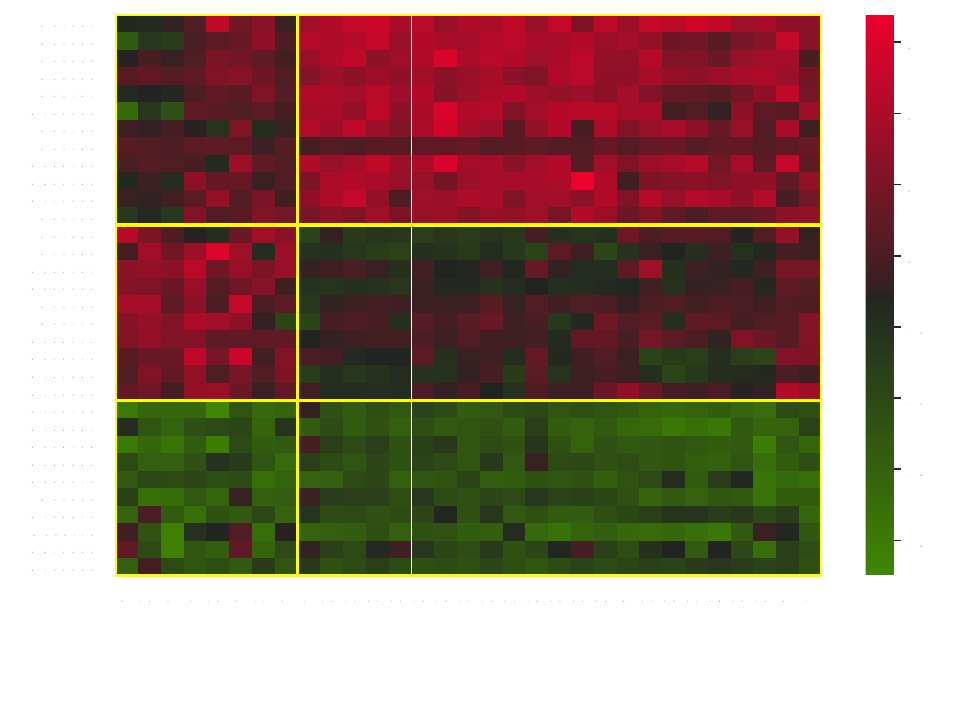}
\caption{Representation of the most similar observations to each archetype. The color represents the expression ratio of each gene for each sample.}
\label{fig:gen:general}
\end{figure}

As can be seen in Fig. \ref{fig:gen:general}, regarding the melanoma samples, if we cluster the data using the location of the maximum archetypal coefficient $\gamma$ (i.e. the archetype that is most similar to the sample), we obtain two clusters of 8 and 23 samples. Regarding the archetypes of the genes, the first two discriminate the two groups quite well, while the third archetype (or group of genes) is not expressed for any melanoma group.

\begin{figure}[!t]
\centering
\subfloat[Representation of the observations ($\alpha$ matrix) in their archetypal space. Each corner represents an archetype.]{\includegraphics[width=0.3\columnwidth]{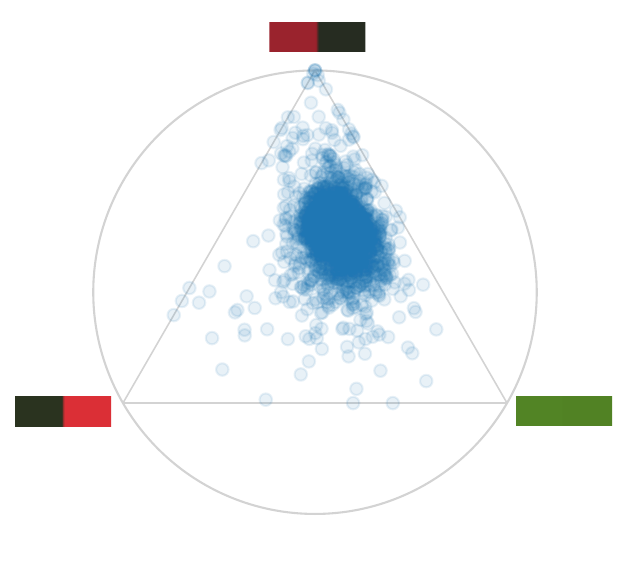}%
\label{fig_first_case}}
\hfil
\subfloat[Representation of the variables ($\gamma$ matrix) in their archetypal space. Each corner represents an archetype. Samples 1, 8, 4 and 7 are colored in red.]{\includegraphics[width=0.27\columnwidth]{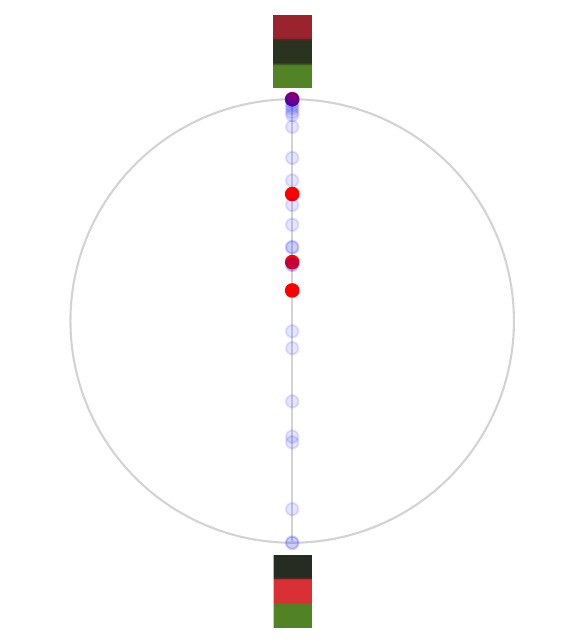}%
\label{fig_second_case}}
\caption{Representations of the archetypal space for Gene expression data.}
\label{fig:gen:partial}
\end{figure}

If we compare our results to those obtained by \cite{Bittner2000},  four samples are classified in different clusters. Samples 1, 4, 7 and 8 belong to the main cluster group with biAA (see Fig. \ref{fig:gen:partial}), unlike results provided by \cite{Bittner2000}. If we compare our results to those obtained by \cite{Rocci2008}, two samples are classified in different clusters, samples 1 and 8. Classification group of samples 4 and 7 is shared with biAA and results by  \cite{Rocci2008}. According to the $\gamma$ coefficients of biAA, samples 4 and 7 are a nearly equal mixture between both archetypes, with the values of the coefficient being 0.4 and 0.6 corresponding to the first and second archetype for sample 4, and 0.45 and 0.55 corresponding to the first and second archetype for sample 7. Therefore, samples 4 and 7 could be in the border between both groups, which could explain the difference in classification by different methods. However, the  $\gamma$ coefficients for sample 8 and 1 are 0.75 (0.25) and 1 (0) for the second (first) archetype, respectively. In other words,  sample 1 (and to lesser extent sample 8) should definitely be in the main group according with biAA. 


\subsection{Text documents}

Another common use for  biclustering is for clustering documents and words. In this case, we have applied biAA to a subset of the 20 Newsgroups collection, set up by \cite{Lang95}. Specifically, we have analyzed three topics: \emph{rec.autos}, \emph{rec.sport.hockey} and \emph{talk.politics.guns}.

{
Additionally, although our algorithm is designed to identify extreme prototypes rather than clusters, due to the absence of comparable benchmarks, we have conducted comparisons with the following popular biclustering algorithms, which are also considered as the baseline  elsewhere \cite{Cotelo2020}: Louvain Clustering \cite{blondel2008fast}, Spectral Co-clustering \cite{dhillon2001co} and Spectral Biclustering \cite{kluger2003spectral}. For the three methods we have left all the default values and, in those that allow it, we have determined the number of clusters to search.
}

Specific, for each document, the number of times each word is repeated in the document has been stored in a count matrix, where each row represents a document, each column a word, and the values indicate how many times each word is repeated in each document.

In addition, a Tfidf transformer \cite{Lang95} was used to convert a count matrix into a normalized tf or tf-idf representation. Tf stands for term frequency, and tf-idf stands for term frequency multiplied by inverse document frequency. This is a standard term weighting method used in information retrieval, and it is also effective for classifying documents.

We have applied biarchetype analysis to this normalized matrix, obtaining three archetypal profiles for documents and three archetypal profiles for words.

\begin{figure}[h!]
\centering
\includegraphics[width=0.3\columnwidth]{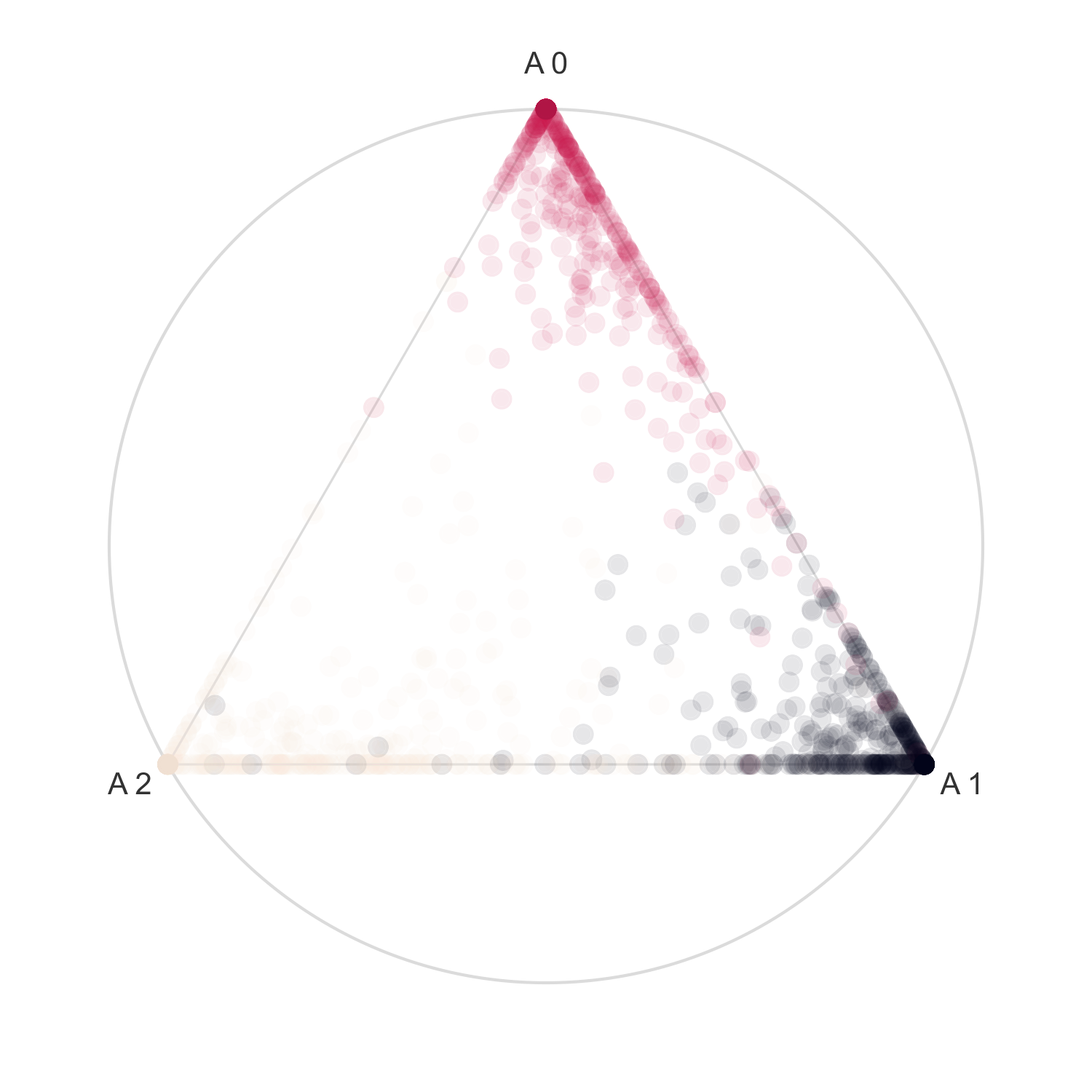}
\caption{Representation of the documents in their archetypal space ($\alpha$ values). The color represents the category of the document.}
\label{fig:doc:doc}
\end{figure}

In Fig. \ref{fig:doc:doc}, it is clear that the three archetypes discriminate the three groups of documents perfectly.

\begin{figure}[h!]
\centering
\includegraphics[width=0.6\columnwidth]{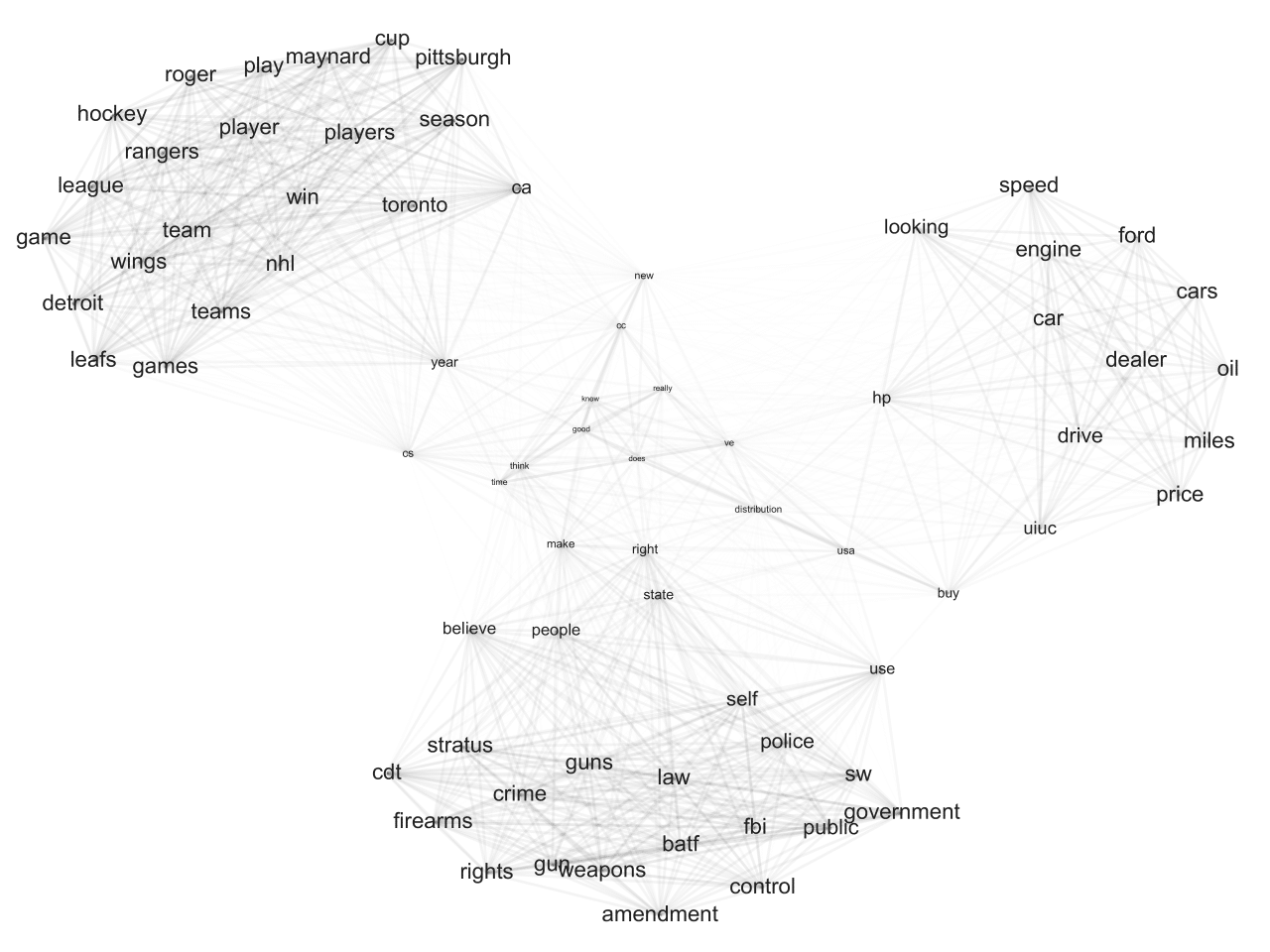}
 \caption{Representation of the most similar words to each archetype (filtered using a threshold over the coefficient matrix). The weight of each edge is the cosine similarity between the two words in their archetypal space. This plot was created using the \emph{networkx} Python package.}
\label{fig:doc:word}
\end{figure}

Regarding the words, in Fig. \ref{fig:doc:word} the words are represented as a graph. The weight of each edge represents the similarity of the words in terms of the archetypes (the gamma coefficients). The graph weights (i.e. the gamma coefficients) split the words into three groups, where the words within each group are related to one of the selected topics.

\begin{figure}[ht!]
    \centering
    \subfloat[PCA and prototypes obtained in the \emph{documents} part.]{
        \includegraphics[width=0.25\columnwidth]{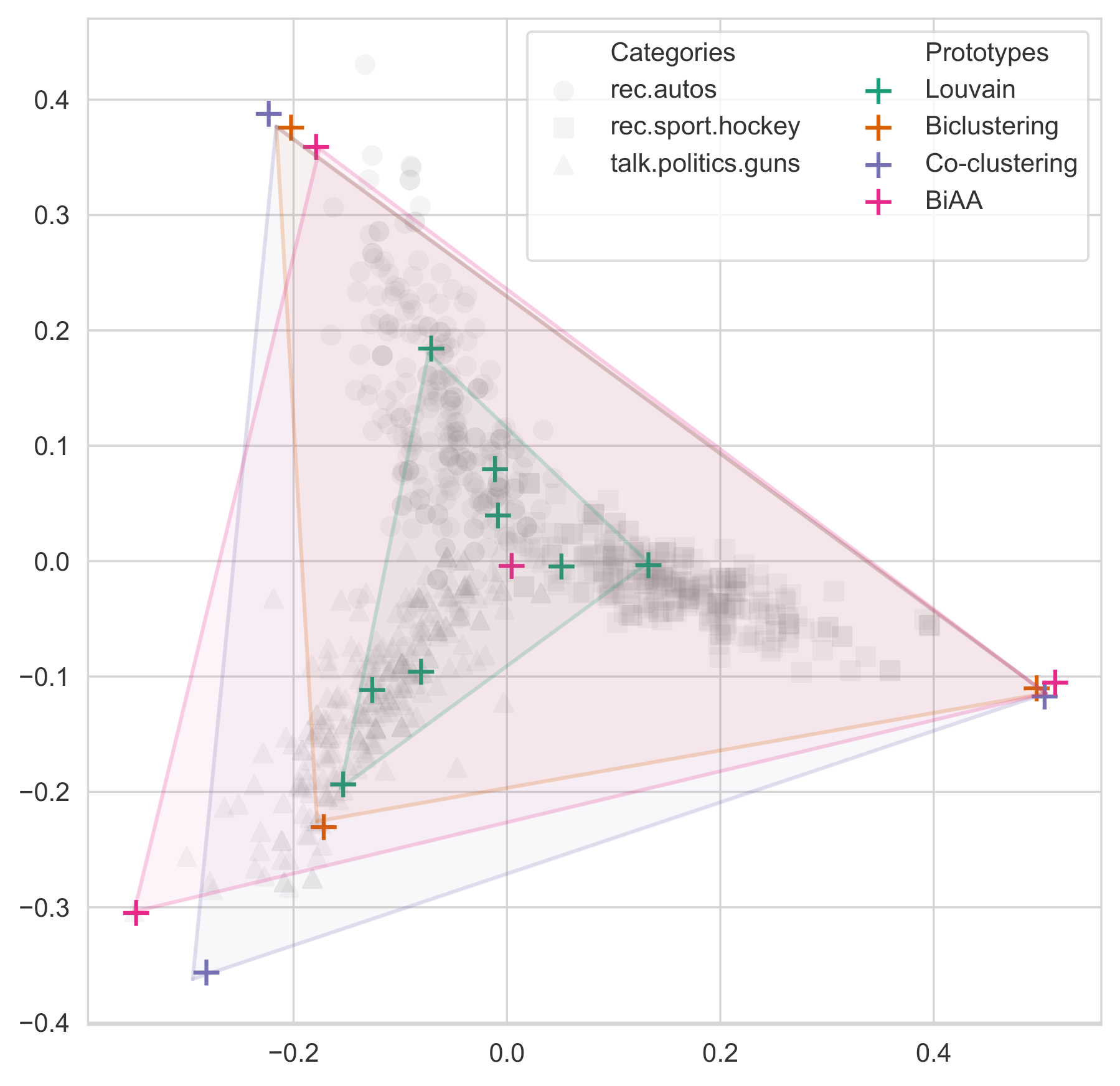}%
        \label{pca_documents}
    }
    \hfil
    \subfloat[PCA and prototypes obtained in the \emph{words} part.]{
        \includegraphics[width=0.25\columnwidth]{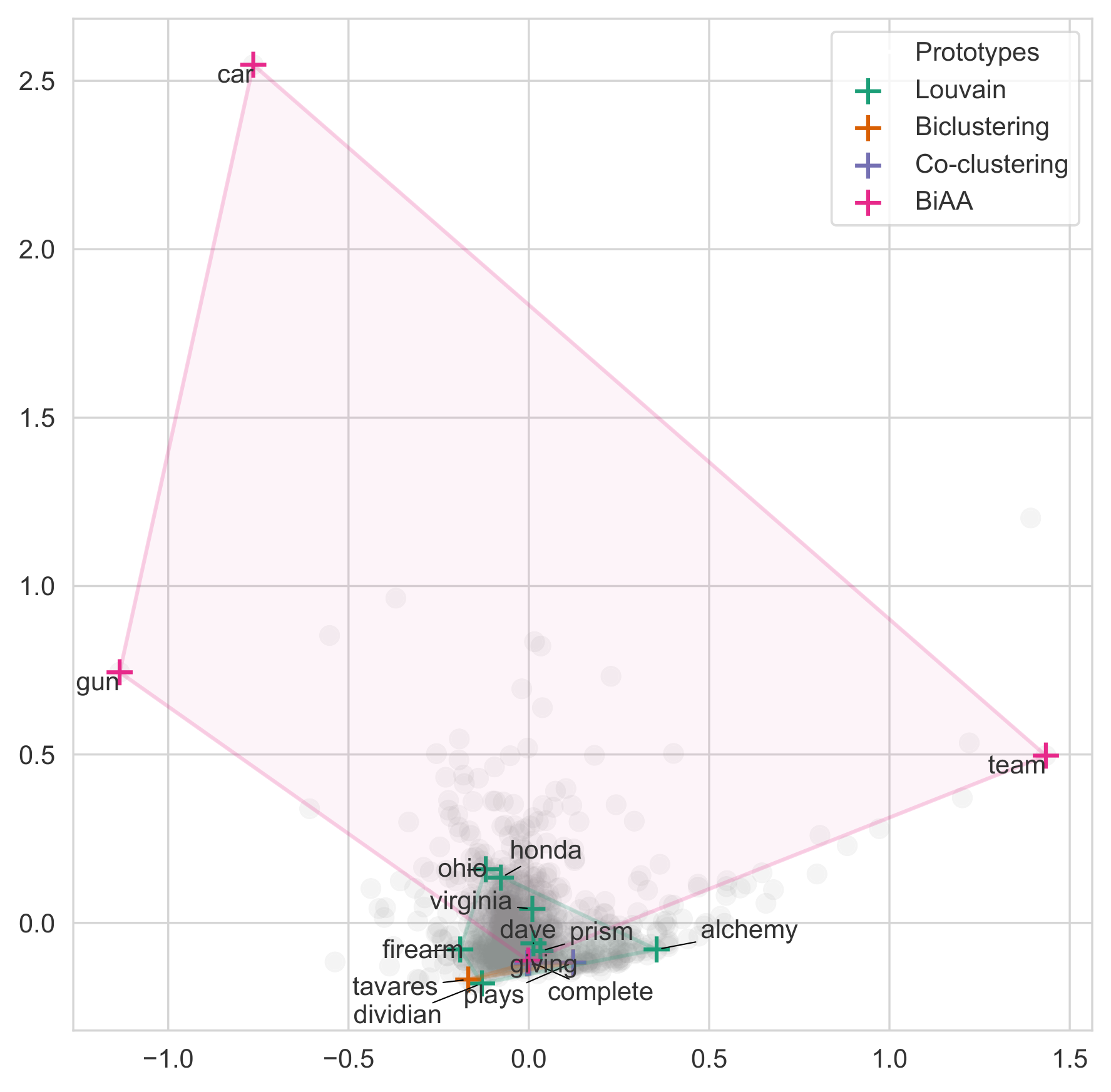}%
        \label{pca_words}
    }     

    \caption{First 2 components of PCA along with the prototypes of the dataset discovered by multiple clustering methods. The colored areas represent the convex hulls of the prototypes for each method.}
    \label{fig:pca}
\end{figure}

{
Upon examining the results of biAA, Figure~\ref{fig:pca} displays the prototypes identified by each method. To facilitate their representation, Principal Component Analysis (PCA) was applied to the data for both rows and columns (the transposed matrix), and the first two components were plotted. It is observable that when analyzing the matrix by rows, which in this example correspond to documents, both  biAA and Spectral Clustering methods yield quite extreme prototypes. However, when the dataset is analyzed from the perspective of words, only  biAA identifies extreme prototypes. Specifically, it identifies the words \emph{car}, \emph{team} and \emph{gun} which correspond to highly archetypal words for groups \emph{rec.autos}, \emph{rec.sport.hockey} and \emph{talk.politics.guns} respectively.
}

\begin{figure}[ht!]
    \centering
    \subfloat[{Biarchetype Analysis.}]{
        \includegraphics[height=1.75cm]{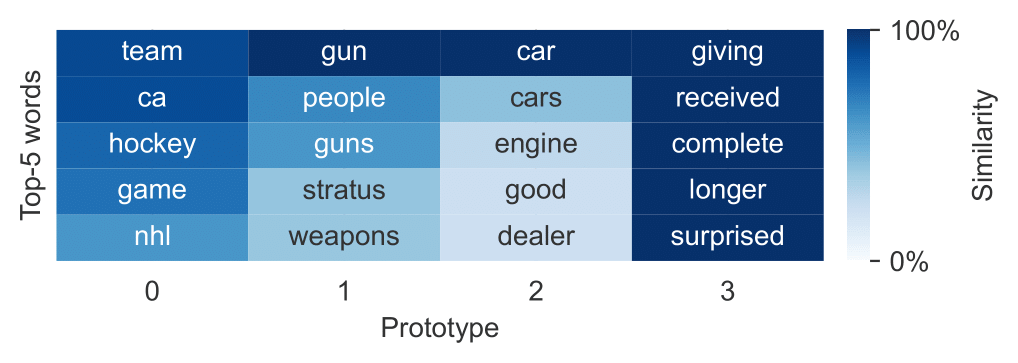}%
        \label{top_biaa}
    }
    \vfil
    \subfloat[{Louvain method.}]{
        \includegraphics[height=1.75cm]{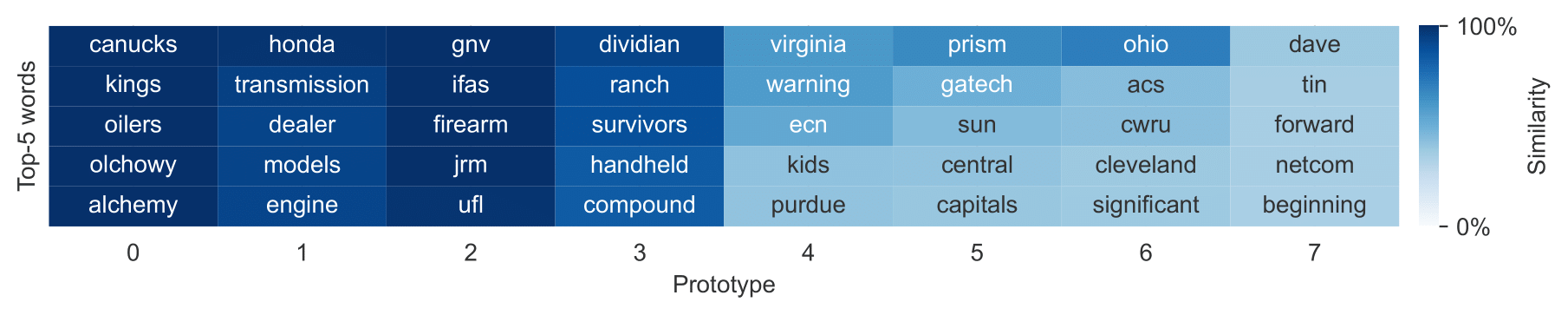}%
        \label{top_louvain}
    }     
    \caption{Top-5  most similar words to each prototype for different algorithms.}
    \label{fig:topwords}
\end{figure}

{
Finally, building on the previous observations, Figure~\ref{fig:topwords} includes the five most similar words  to each prototype for methods that allow for mixed membership, i.e. biAA and the Louvain method. It is evident that the biAA identifies words that are more archetypical compared to those identified by the Louvain method. Specifically, the first three prototypes discovered by biAA can be clearly associated with the three groups of documents present in the dataset.
}

\begin{table}[ht]community detection 
    \caption{RSS for text documents  and community detection example. }
    \label{tab:text-ev}
    \centering
    
\begin{tabular}{|lccc|}
\hline
       Method & RSS (text) & RSS (community) &  Fuzzy \\
\hline
         biAA &             1605.58 & 1064.4&   True \\
      Louvain &             1620.85 & 1268.24 & True \\
 Biclustering &             1664.42 & 1452.03& False \\
Co-clustering &             1666.97 & 1663.77& False \\
\hline
\end{tabular}
\end{table}

{Table \ref{tab:text-ev} compiles RSS for all the methods. biAA provides the lowest RSS.}

\subsection{Community detection}

Finally, we have also applied biAA to detect communities within the company Enron. For that, we have studied the data set described in \cite{Enron}, which contains a collection of emails between the company's employees.

We have created an adjacency matrix between employees, containing 1 if one employee has emailed another  or 0 if the first one has never sent an email to the second one.

\begin{figure}[h!]
\centering
\includegraphics[width=0.5\columnwidth]{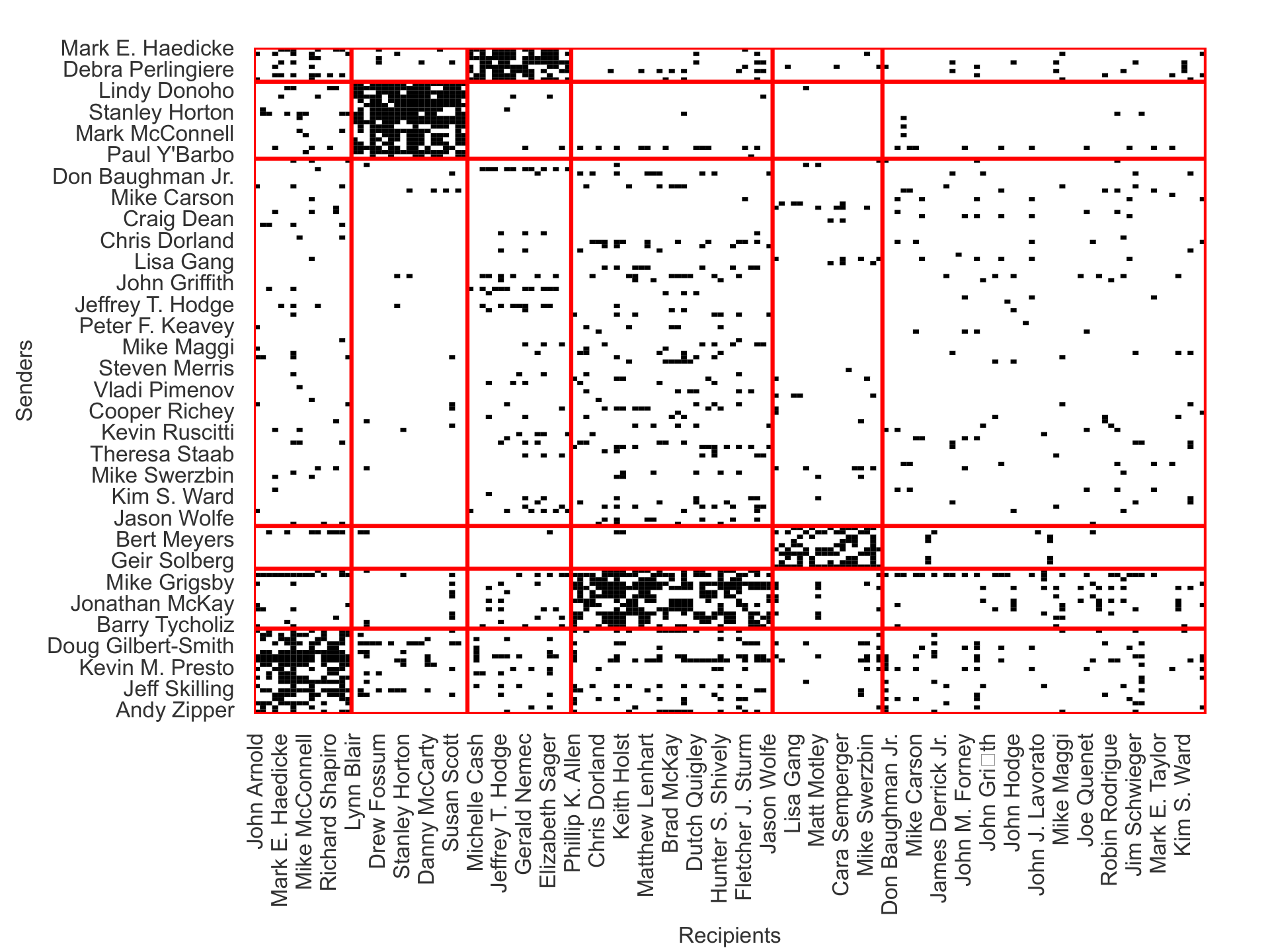}
\caption{The adjacency matrix ordered according to the archetypes obtained with biAA.}
\label{fig:mail:general}
\end{figure}

After applying biAA  with $k=6$ and $c=6$ to this adjacency matrix, we obtained the results in Fig. \ref{fig:mail:general}. In the `senders' part, the group $Z_3$  could be omitted (represents employees who haven't sent emails to a specific group). The same occurs with the group $Z_6$ in the `recipients' part of the adjacency matrix.

\begin{figure}[h!]
\centering
\includegraphics[width=0.6\columnwidth]{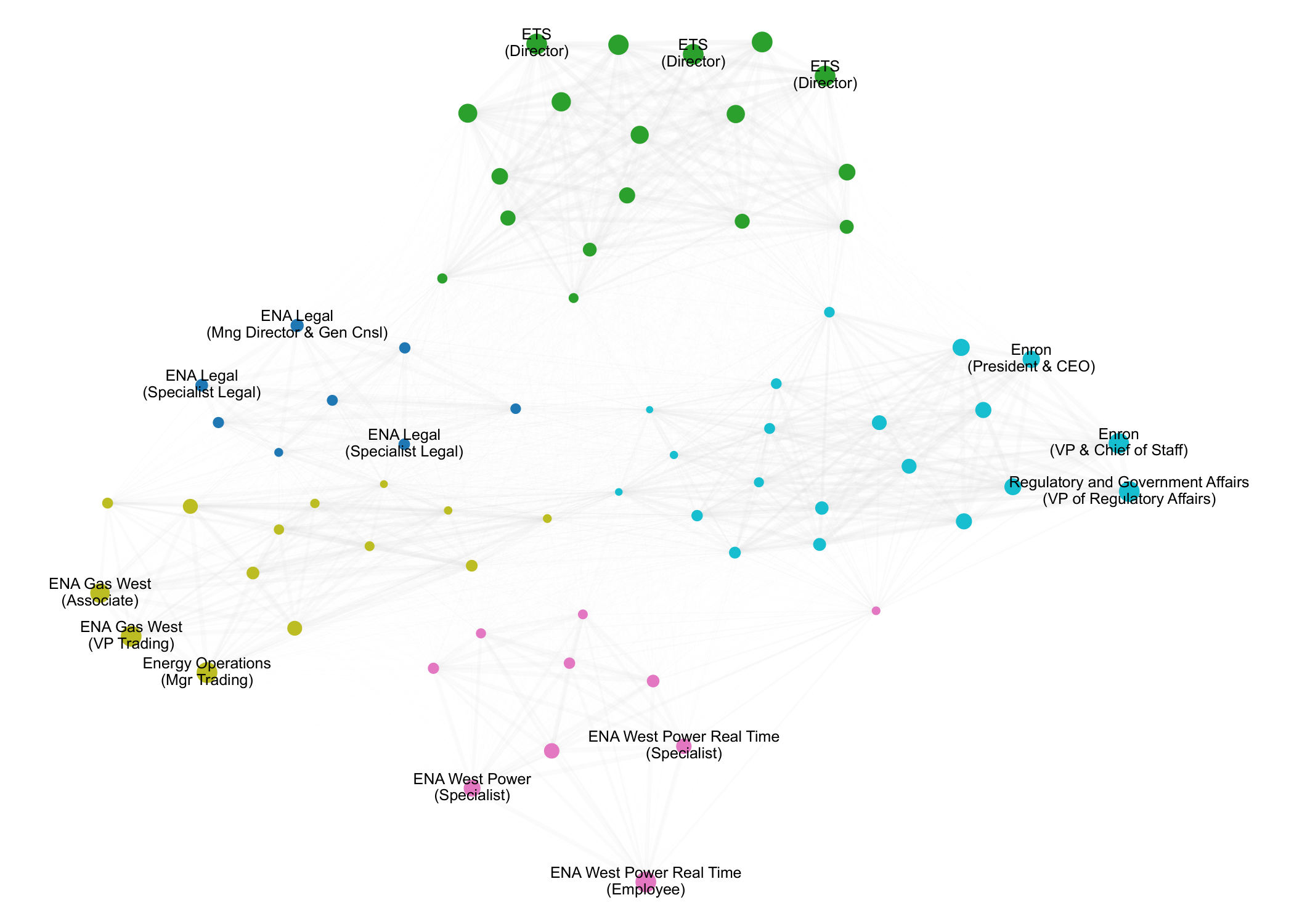}
\caption{Representation of employees from the point of view of who they send emails to.  The weight of each node is computed as in Fig. \ref{fig:doc:word}. The size of each employee is proportional to how similar it is to its closest archetype and the color of each one is determined by the closest archetype.}
\label{fig:mail:senders}
\end{figure}

If we analyze the employees from the point of view of who they send emails to, we obtain the results shown in Fig. \ref{fig:mail:senders}, in which we have removed $Z_3$-like employees as they do not exchange emails with a specific group. As can be seen, the dark blue cluster represents the Legal department, the green one, the ETS department; the olive green one, the Gas/Energy department; the pink one, the West Power departments; and the blue one, Enron's top management.

{

\begin{figure}[ht!]
    \centering
    \subfloat[{PCA and prototypes obtained in the \emph{senders} part.}]{
        \includegraphics[width=0.25\columnwidth]{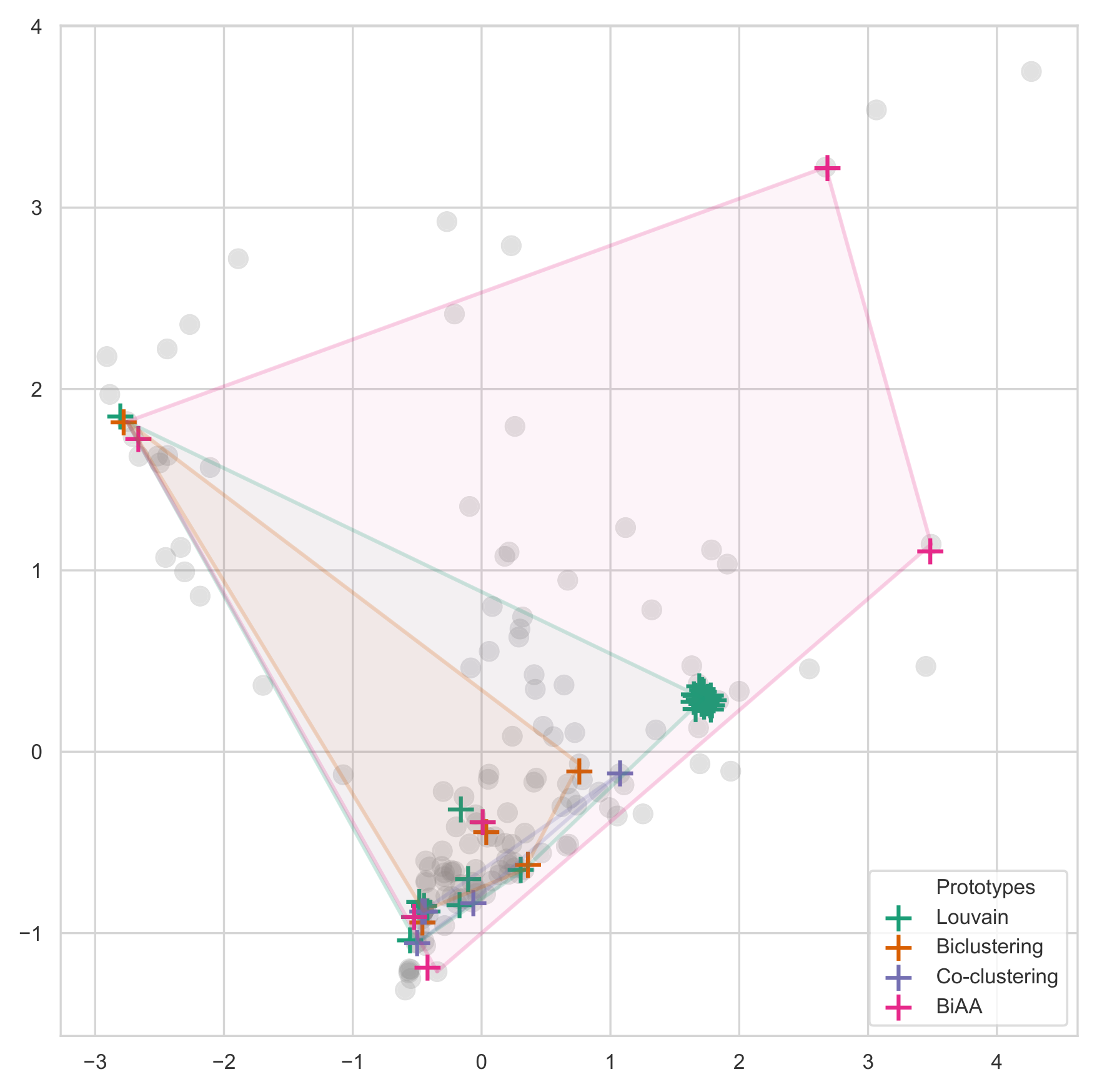}%
        \label{pca-enron-1}
    }
    \hfil
    \subfloat[{PCA and prototypes obtained in the \emph{recipients} part.}]{
        \includegraphics[width=0.25\columnwidth]{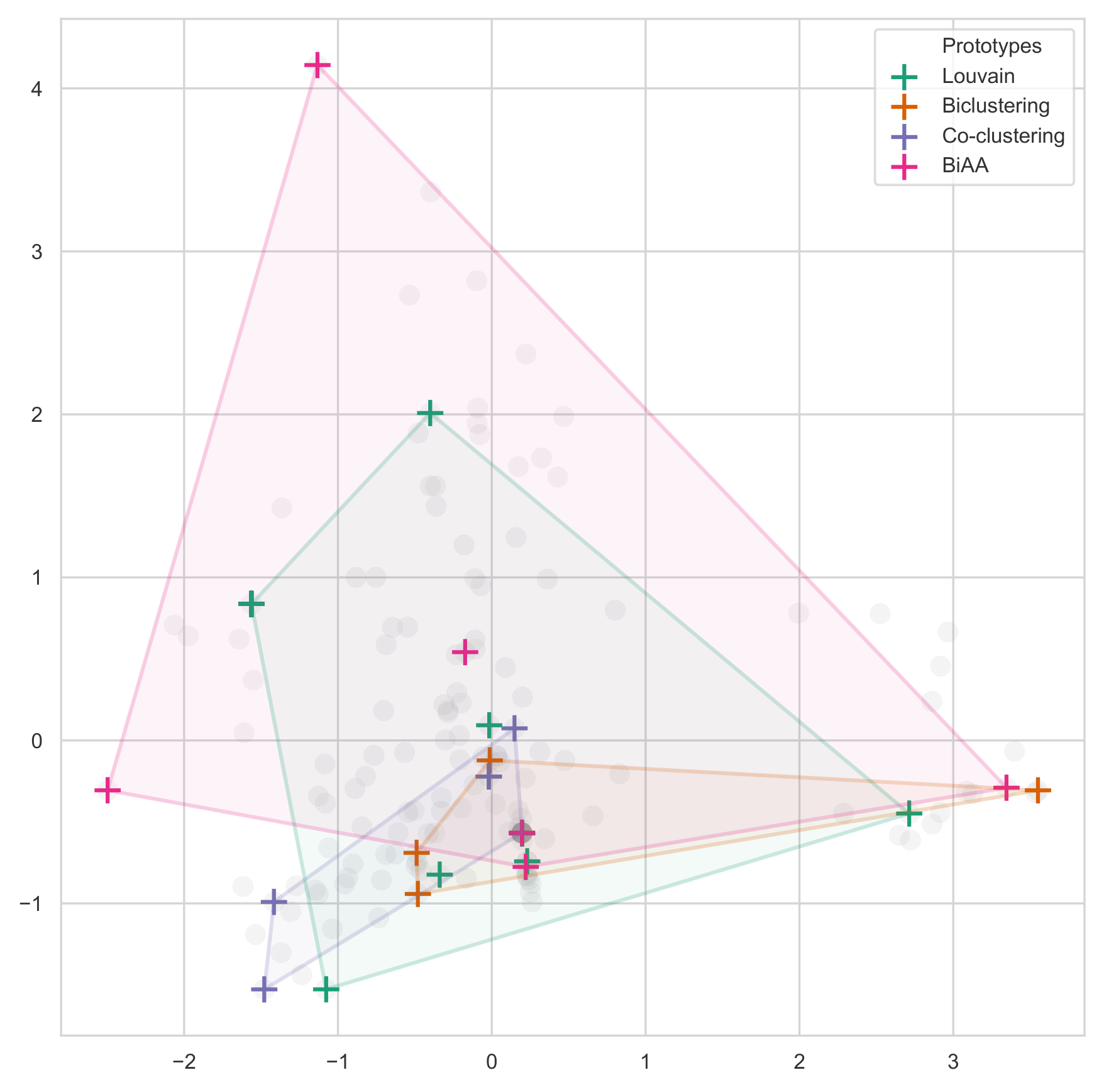}%
        \label{pca-enron-2}
    }     

    \caption{{ First 2 components of PCA along with the prototypes of the dataset discovered by multiple clustering methods. The colored areas represent the convex hulls generated by the prototypes of each method.}}
    \label{fig:pca-enron}
\end{figure}


Here, in Figure~\ref{fig:pca-enron} the same procedure as in the previous problem has been applied. It is clear that in both cases, for both the senders and the recipients of the emails, the convex hull of the prototypes identified by biAA covers the largest area. This could serve as a measure of how extreme (or distant from each other) the prototypes are. Therefore, based on this metric, it is evident that  biAA reveals the most extreme prototypes.
}

{As before, biAA also provides the lowest RSS for this example.}

\section{Conclusion \label{conclusiones}}
In this work, we have proposed a new unsupervised machine learning technique: biarchetype analysis.  We have  compared the results of biAA and biclustering in an illustrative example, showing not only the greater interpretability provided by biAA, but also the greater coherence of the results. We have {also} seen its usefulness in several problems in different fields, {where more distinct aspects are extracted with biAA than using several biclustering methods}.

 BiAA has been defined for continuous data. In future work, it could be extended to other kinds of data, such as functional data, to which AA was also extended \cite{Epifanio2016}. Note that biclustering analysis of time series is used in many fields such as neuroscience \cite{castanho2022biclustering} and engineering \cite{w14121954}; therefore, biAA could also be used for the same problems. Biarchetypoid analysis could also be introduced in the same way that archetypoid analysis was defined \cite{Vinue15}, where biarchetypes are not determined by mixtures of observations and features, but by concrete elements of the data set. Just as archetype analysis is sensitive to outliers, biAA is too. Robust biAA could be defined in the same way as robust AA was \cite{Moliner2018a}. Likewise, biAA for missing data could be defined as it was for AA \cite{doi:10.1080/00031305.2018.1545700}, and it could be used in recommender systems to find profiles of users and products, for instance. Another line of future work would be to apply biAA to different fields where biclustering analysis is applied, and to study more computational methods to calculate biAA, especially for big data. {Furthermore,} biAA could also be easily extended to high dimensions in a similar way to the decomposition proposed in \cite{tucker1966some}. {Finally, non-linear biAA could be proposed by using deep learning, based on the works on deep AA by  \cite{van2019finding} and \cite{keller2021learning}.}


%

\section*{Acknowledgments}

{The authors would like to thank Francesca Martella for providing them with gene expression data.}

This research was partially supported by the Spanish Ministry of Universities (FPU grant FPU20/01825), Spanish Ministry of Science and Innovation (PID2022-141699NB-I00, PID2020-118763GA-I00 and PID2020-115930GA-I00) and UJI-B2020-22 and TRANSUJI/2023/6 from Universitat Jaume I, Spain.

\bibliographystyle{apalike}

\bibliography{biaa}

\appendix
\section{Proof of Propositions 1 and 2 \label{ape}}

Proof of {\it Proposition 1}. 

Let us denote ${\bf H}_{n\times m}=\alpha_{n \times k} {\bf Z}_{k \times c} \gamma_{c \times m}$ and $ {\bf R}_{n \times c} =  \alpha_{n \times k} {\bf Z}_{k \times c}$;
then, each ${\bf r}_j^d $ $(j=1,\dots, n)$ belongs to the convex hull $C_Z^d$ of ${\bf z}_i^d$ $(i=1,\dots, k)$. Moreover, since $ {\bf H}_{n \times m} = {\bf R}_{n \times c} {\gamma}_{c \times m}$, each vector ${\bf h}_j^f$ $(j=1,\dots, m)$ belongs to the convex hull $C_R^f$ of the vectors  ${\bf r}_i^f$ $(i=1,\dots, c)$.

Having fixed ${\bf Z}_{k \times c}$; that is, having fixed ${\bf z}_j^d$ $(j=1,\dots, k)$,  $RSS =  \left \Arrowvert {\bf X}_{n \times m} - {\bf H}_{n \times m}\right \Arrowvert^2$ is minimized with respect to $\alpha's$ and $\gamma's$ by choosing $\forall i=1,\dots,n$, ${\bf h}_i^d$ to be the point in the convex set defined by $C_R^f$ and $C_Z^d$ that is closest to ${\bf x}_i^d$.

Suppose without loss of generality that ${\bf z}_1^d$ is strictly interior to $C_V^d$; then choose $t$ such that ${\bf z}(t)={\bf z}_j^d +t ({\bf z}_1^d - {\bf z}_j^d)$ is on the boundary of $C_V^d$. Since the matrix ${\bf \theta}_{m \times c}$ is considered fixed, the convex hull $C$ of ${\bf z}(t), {\bf z}_2^d,\dots,{\bf z}_k^d$ contains the convex hull $C_Z^d$ of ${\bf z}_1^d, {\bf z}_2^d,\dots,{\bf z}_k^d$; therefore, we obtain a larger set over which to minimize $RSS =  \left \Arrowvert {\bf X}_{n \times m} - {\bf H}_{n \times m}\right \Arrowvert^2$. 
\medskip

\noindent Proof of {\it Proposition 2}. 

Now, let us denote  $ {\bf S}_{k \times m} =  {\bf Z}_{k \times c} \gamma_{c \times m}$;
then, each ${\bf s}_j^f $ $(j=1,\dots, m)$ belongs to the convex hull $C_Z^f$ of ${\bf z}_i^f$ $(i=1,\dots, c)$. Moreover, since $ {\bf H}_{n \times m} = \alpha_{n \times k}{\bf S}_{k \times m}$, each vector ${\bf h}_j^d$ $(j=1,\dots, n)$ belongs to the convex hull $C_S^d$ of the vectors  ${\bf s}_i^d$ $(i=1,\dots, k)$.

As before, having fixed ${\bf Z}_{k \times c}$; that is, having fixed ${\bf z}_j^d$ $(j=1,\dots, k)$,  $RSS =  \left \Arrowvert {\bf X}_{n \times m} - {\bf H}_{n \times m}\right \Arrowvert^2$ is minimized with respect to $\alpha's$ and $\gamma's$ by choosing $\forall i=1,\dots,n$, ${\bf h}_i^f$ to be the point in the convex set defined by $C_S^d$ and $C_Z^f$ that is closest to ${\bf x}_i^d$.

Suppose without loss of generality that ${\bf z}_1^f$ is strictly interior to $C_Y^f$; then choose $t$ such that ${\bf z}(t)={\bf z}_j^f +t ({\bf z}_1^f - {\bf z}_j^f)$ is on the boundary of $C_Y^f$. Since the matrix ${\bf \beta}_{k \times n}$ is considered fixed, the convex hull $C$ of ${\bf z}(t), {\bf z}_2^f,\dots,{\bf z}_c^f$ contains the convex hull $C_Y^f$ of ${\bf z}_1^f, {\bf z}_2^f,\dots,{\bf z}_c^f$; therefore we obtain a larger set over which to minimize $RSS =  \left \Arrowvert {\bf X}_{n \times m} - {\bf H}_{n \times m}\right \Arrowvert^2$.

\end{document}